\documentclass[aps,prf,reprint,unsortedaddress,onecolumn]{revtex4-2}

\usepackage{natbib}    
\usepackage{amsmath, amsthm, amssymb}   
\usepackage{graphicx}                   
\usepackage{url}
\usepackage{ulem, CJKulem}              
\usepackage{booktabs}                   
\usepackage{multirow}                   
\usepackage{diagbox}                    
\usepackage{array}                      
\usepackage{longtable}                  
\usepackage{paralist}                  
\usepackage{hyperref}
\usepackage{tikz}
\usetikzlibrary{shapes,arrows}
\usepackage{verbatim}
\usepackage[none]{hyphenat}
\usepackage{natbib}
\usepackage{CJKutf8}
\usepackage[utf8]{inputenc}
\usepackage[percent]{overpic}
\usepackage{subfigure}
\usepackage{orcidlink}
\UseRawInputEncoding

\begin{document}
\title{Weakly coupled fluid--structure interaction between wall-bounded turbulent flows and defect-embedded phononic subsurfaces}

\author{Ching-Te Lin\footnote{Authors contributed equally to this work.\label{fn1}}\orcidlink{0000-0001-5919-8718}}
\affiliation{Department of Mechanical and Civil Engineering, California Institute of Technology, Pasadena, CA 91125, USA.}
\author{Vinod Ramakrishnan\footref{fn1}\orcidlink{0000-0002-5588-876X}}
\affiliation{The Grainger College of Engineering, Department of Mechanical Science and Engineering, University of Illinois at Urbana-Champaign, Urbana, IL 61801, USA}
\author{Andres Goza\orcidlink{0000-0002-9372-7713}}
\affiliation{The Grainger College of Engineering, Department of Aerospace Science and Engineering, University of Illinois at Urbana-Champaign, Urbana, IL 61801, USA}
\author{Kathryn H. Matlack\orcidlink{0000-0001-7387-2414}}
\affiliation{The Grainger College of Engineering, Department of Mechanical Science and Engineering, University of Illinois at Urbana-Champaign, Urbana, IL 61801, USA}

\author{H. Jane Bae\orcidlink{0000-0001-6789-6209}}
\affiliation{Lynn Booth and Kent Kresa Department of Aerospace, California Institute of Technology, Pasadena, CA 91125, USA.}

\begin{abstract}

We investigate the interaction between wall-bounded turbulence and defect-embedded phononic subsurface (D-Psub) using a weakly coupled fluid--structure framework, in which the flow and structure are advanced sequentially without sub-iterations. The D-Psub subsurface is modeled as a dynamic wall with a resonance introduced via a localized structural defect, driven by spatially averaged wall-pressure fluctuations from a turbulent channel flow. This configuration enables a controlled study of how a narrow-band structural response interacts with the broadband forcing of near-wall turbulence.
Despite broadband turbulent forcing, the D-Psub exhibits a narrow-band response that modifies near-wall dynamics, with representative cases showing suppression of velocity fluctuations, increased coherence of streamwise streaks, and a measurable reduction in turbulent drag. Crucially, the coupled system displays behavior that cannot be replicated by prescribed wall motion: the dominant oscillation frequency shifts away from the designed resonance due to fluid--structure interaction. Additionally, the phase between panels is shown to be governed by the convection of turbulent structures.
These results reveal a mechanism by which phononic subsurfaces filter and reorganize turbulent energy through frequency-selective coupling, distinct from conventional compliant or actively forced walls. The findings provide a physical basis for designing passive resonant surfaces that exploit turbulence--structure coupling for flow control.

\end{abstract}


\maketitle

\section{Introduction}\label{sec:Introduction}

Turbulent flows interacting with solid boundaries are governed by multiscale, nonlinear dynamics that give rise to complex near-wall structures. Modifying these structures through surface-based strategies has long been of interest, both for practical objectives such as drag reduction and noise attenuation, and for gaining physical insight into turbulence–wall interactions \citep{bushnell1989turbulence}. Classical approaches rely on static surface modifications, including riblets \citep{Garcia_11, choi_1993, liu_12}, roughness manipulation \citep{Chung_2015}, and superhydrophobic coatings \citep{Gose_2018, Rastegari_Akhavan_2015, Rastegari_Akhavan_2019}, which alter near-wall turbulence through fixed geometric features. These methods have demonstrated measurable success when their characteristic length scales are appropriately matched to the flow \citep{choi_1993}. However, this effectiveness is inherently tied to specific flow conditions, such as Reynolds number, which determine the relevant turbulent scales. As these conditions change, the optimal interaction scales shift, while static surfaces remain fixed. This lack of adaptability fundamentally limits their robustness and motivates the search for surfaces capable of dynamically responding to the flow.

In wall-bounded turbulence, momentum transfer is governed by coherent near-wall structures, including streamwise streaks, quasi-streamwise vortices, and their self-sustaining regeneration cycle, which are responsible for the majority of turbulent production and Reynolds stress near the wall \citep{Hamilton_1995,Waleffe_97,Jimenez_2018}. These structures exhibit characteristic spatial and temporal scales linked to viscous and shear-based quantities, yet they coexist with a broad range of fluctuations spanning multiple scales\citep{Sillero_14}. Consequently, the wall-pressure signature generated by turbulence is inherently broadband and stochastic\citep{Choi_and_Moin_1990,Yang_Yang_2022}. This combination of organized structures embedded within broadband forcing presents a fundamental challenge for surface-based interaction: a surface must be able to respond to a wide spectrum of fluctuations while selectively coupling to dynamically relevant motions. Achieving such selective interaction in a multiscale, time-dependent environment remains a central difficulty in modifying wall-bounded turbulence.

To overcome these limitations, dynamic surfaces that respond to fluid forcing have been explored as an alternative paradigm \citep{Bushnel_77,Duncan_1986,kireiko1990}. Compliant coatings and other passive dynamic control strategies introduce additional degrees of freedom \citep{Kramer60,Carpenter_Garrad_1985,Carpenter_Garrad_1986,Wang05,Wang06}, allowing the wall to deform in response to fluid pressure and thereby interact with the flow in a time-dependent manner. Early theoretical and experimental studies demonstrated that compliant surfaces can delay laminar-to-turbulent transition \citep{Hussein15, Wang_Koley_Katz_2020, Fabbiane_2025} and, under carefully tuned conditions, reduce drag in turbulent flows \citep{Endo01012002, Fukagata01012008, Kim_Choi_2014, Jozsa_19}. However, sustained and robust performance in fully developed turbulence has remained elusive. The response of conventional compliant surfaces is typically governed by a small set of bulk material parameters—such as stiffness, damping, and density—which, when modeled, are often represented using lumped mass--spring--damper systems \citep{Fukagata_02, Esteghamatian_Katz_Zaki_2022}. These formulations produce a broadband mechanical response, meaning the surface reacts to a wide range of flow scales simultaneously, without the ability to selectively couple with dynamically relevant structures. As a result, the interaction is highly sensitive to the specific material properties and flow conditions: small deviations from an optimal parameter regime can lead to performance degradation or even drag increase, and the underlying physical mechanisms remain difficult to isolate and control.

A key limitation of existing compliant-wall models is their reliance on low-dimensional representations, such as lumped mass--spring--damper systems, which restrict both the richness of the achievable wall response and the interpretability of the resulting flow modifications. While such models have proven useful for identifying broad trends, their response is typically governed by a small set of parameters and remains inherently broadband, lacking intrinsic mechanisms for frequency-selective or spatially structured interaction. As a result, they do not naturally enable targeted coupling with specific turbulent scales. Consequently, a central unresolved question in the field is how to design surfaces that interact with turbulence in a scale-selective, dynamically robust manner—coupling meaningfully with a chosen subset of the turbulent spectrum while remaining insensitive to others.

Recent advances in metamaterials \citep{Avallone2026} offer a fundamentally new approach to this problem. Among these, phononic materials are architected structures whose dynamic response can be tailored through their internal geometry, enabling frequency-selective behavior via band gaps and spatially-localized resonances \citep{Kushwaha_93, ZhangAFM2025,PatilAM2022,RamakrishnanJASA2023,NematNasserAIP2011}. Band gaps are frequency ranges in which wave propagation through the material is suppressed; spatially-localized modes can arise either by truncating the infinitely periodic unit cell (forming ``truncation resonances") or from a local perturbation to the periodic structure (forming ``defect resonances"). The resulting spatially-localized resonance within the band gap produces a narrow-band response at a tunable frequency that depends on the geometry and phononic material parameters. The combination of broadband attenuation punctuated by sharp, geometry-controlled resonances stands in fundamental contrast to conventional compliant surfaces and offers a new mechanism for selective turbulence--wall coupling. While related concepts have been explored in the context of aeroacoustics \citep{SchmidtJAP2025}, flow instabilities \citep{Hussein15, MichelisPoF2023}, and laminar boundary layer control \citep{WILLEY23, NavarroMatter2025}, their interaction with fully developed turbulent boundary layers has not been systematically investigated.

Phononic subsurfaces are therefore well-suited to address the scale-selectivity challenge identified above. By designing the defect properties of a phononic subsurface, it is possible to embed a narrow-band resonance at a frequency corresponding to dynamically significant turbulent scales, such as the characteristic frequency of near-wall streaks or the peak of the wall-pressure spectrum, while attenuating the response at other frequencies. The resulting surface response is not merely a passive filter but a potential source of organized feedback onto the flow, with consequences for near-wall structure and turbulent transport that remain largely unexplored.

In this work, we investigate the interaction between turbulent channel flow and defect-embedded phononic subsurfaces (D-Psubs) developed by \citet{RamakrishnanJSV2025} within a weakly coupled fluid--structure framework, in which the flow and structure are advanced sequentially without sub-iterations. In this formulation, the structure responds to the flow without modifying the instantaneous fluid solve, introducing a one-way lagged coupling that isolates the leading-order interaction mechanisms. The D-Psub is modeled as a dynamic wall with a defect-induced resonance, driven by wall-pressure fluctuations from the turbulent flow. This configuration enables a controlled study of how a narrow-band structural response interacts with the broadband forcing of near-wall turbulence.

The objectives of this study are to (i) characterize the coupled dynamics between turbulence and phononic subsurfaces, (ii) identify how frequency-selective structural response modifies near-wall turbulence, and (iii) elucidate the physical mechanisms governing this interaction. By systematically varying the defect properties of the D-Psub, we explore a range of resonant behaviors and their impact on the flow. The results reveal emergent coupling phenomena — including shifts in structural response frequency and convection-governed phase relationships between panels — highlighting the fundamentally bidirectional nature of the turbulence--phononic subsurface interaction.

The remainder of the paper is organized as follows. The modeling framework is described in Section~\ref{sec:Simulations}. The selection of D-Psub parameters is presented in Section~\ref{sec:RPM_selection}. The effects on turbulent flow statistics and the coupled phononic subsurface dynamics are discussed in Sections~\ref{sec: RPM dynamics} and Section~\ref{sec: turbulent stats}, respectively. Finally, conclusions are given in Section~\ref{sec: conclusion}.

\section{Modeling Framework} \label{sec:Simulations}
\begin{figure}
    \centering
    \includegraphics[width=\textwidth]{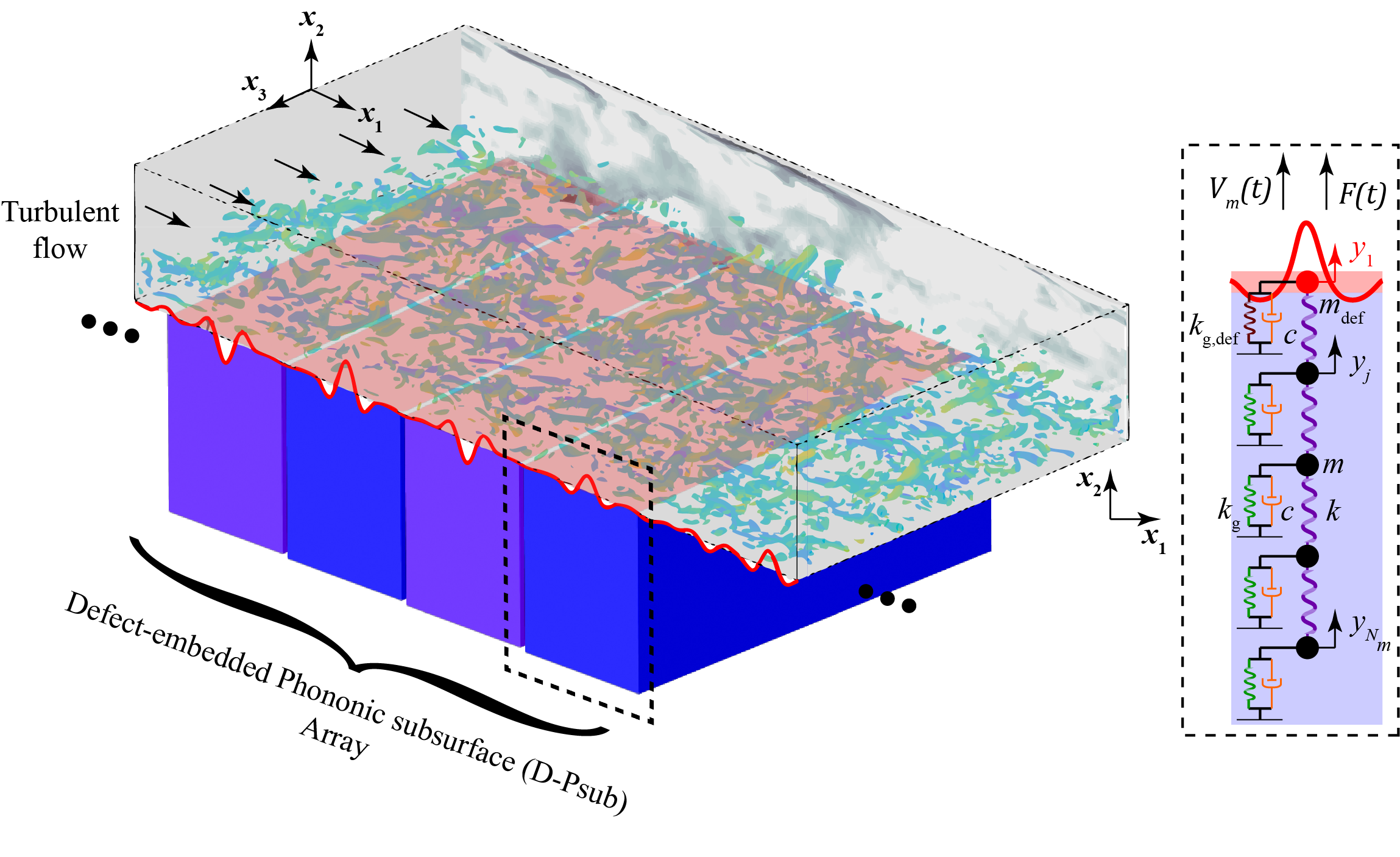}
    \caption{Schematic of the weakly coupled fluid--D-Psub interaction framework. The turbulent channel flow provides an unsteady wall-pressure load, which is surface-averaged over each D-Psub control panel and applied as a temporal forcing to the reduced-order D-Psub model. The resulting wall-normal velocity response is mapped back to the fluid domain as a spatially localized Gaussian transpiration boundary condition, enforcing a zero-net-flux constraint at the wall.}
    \label{fig:schematic}
\end{figure}

We consider the weakly-coupled dynamics of a turbulent channel flow over a phononic subsurface, where the phononic subsurface is fully embedded within the bottom wall such that there is an impermeable FSI interface (Fig.~\ref{fig:schematic}). The phononic subsurface is represented as a chain of mass--spring--damper model that only deforms in the wall-normal direction, and the resulting wall motion is approximated as a blowing and suction profile at the nominal wall location.

\subsection{Reduced-order model for phononic subsurface dynamics} \label{sec:RPM}

The dynamics of the phononic subsurface subjected to unsteady normal loading are approximated using a reduced-order linear system represented by a defect-embedded phononic crystal (PnC)\citep{RamakrishnanJSV2025}, an architected material that can leverage its structural periodicity to enable wave dispersion and customize the spatio-temporal characteristics of these waves. 

Consider first a finite \textit{defect-free} PnC with $N_m$ degrees of freedom all with the same mass, with elastic grounding. This system admits $N_m$ natural modes characterized by discrete wavenumber--frequency pairs $(\kappa_j,\omega_j)$. Introducing a structural defect by locally modifying the mass and/or stiffness properties breaks the periodicity and shifts one of these modes into the band gap, giving rise to a localized defect resonance at frequency $\omega_\mathrm{def}$. This defect mode is associated with an exponentially decaying spatial structure that concentrates energy near the defect degree of freedom.  To concentrate the structural energy near the fluid--phononic interface, we place the defect in the degree of freedom at the fluid--phononic interface (i.e.~$m_{\mathrm{def}}$ in Fig.~\ref{fig:schematic}).

The dispersion relations of this defect-embedded PnC exhibit two distinct spectral regions in the wavenumber--frequency domain: the pass band (PB), where waves propagate with real wavenumber $\kappa_\mathrm{R}$, and the band gap (BG), where waves are spatially attenuated due to complex wavenumber $\mathrm{i}\kappa_\mathrm{I}$ or $\pi+\mathrm{i}\kappa_\mathrm{I}$, where $\mathrm{i} = \sqrt{-1}$ \citep{Deymier2013}. As a result, disturbances with frequencies in the PB can propagate through the material, whereas those within the BG are exponentially localized and do not propagate into the bulk.

Note that this combination of spectral filtering and spatial localization is particularly advantageous for interaction with turbulent flows, whose forcing is inherently broadband. The band gap suppresses response to off-resonant frequencies, while the defect mode enables a narrow-band, high-amplitude response to a targeted frequency range. Consequently, the phononic subsurface can selectively respond to specific components of the broadband turbulent forcing while attenuating others, providing a mechanism for scale-selective fluid--structure interaction (see \citet{LinAIAASciTech2026} for additional details).

The governing equation of the PnC is given by a linear ordinary differential equation 
\begin{equation}
\label{eq:RPM_equation}
    \mathbf{{M}} \ddot{\mathbf{y}} + \mathbf{{C}} \dot{\mathbf{y}} + \mathbf{{K}} \mathbf{y} = \mathbf{F},
\end{equation}
where $\mathbf{y}=[{y}_1,{y}_2,\cdots,{y}_{N_m}]^\mathrm{T}$ represents the displacement vector of an PnC with $N_m$ masses, ${F}$ represents the force exerted on the phononic subsurface, and
\begin{equation*}
    \mathbf{{M}} = \begin{bmatrix} 
        {m}_\mathrm{def} & 0 & \cdots & \cdots & 0 \\ 
        0 & {m} & \cdots & \cdots & 0 \\ 
        \vdots & \vdots & \ddots &  & \vdots \\ 
        \vdots & \vdots &  & \ddots & \vdots \\ 
        0 & 0 & \cdots & \cdots & {m}   
    \end{bmatrix}_{{N_m}\times{N_m}}, \quad
    \mathbf{{C}} = {c} \mathbf{I}_{{N_m}\times{N_m}}, \quad
    \mathbf{{K}} = \begin{bmatrix} 
        {k}+{k}_\mathrm{g,def} & -{k} & 0 & \cdots & 0 \\ 
       -{k} & 2{k}+{k}_\mathrm{g} & -{k} & \cdots & 0 \\ 
        0 & -{k} & \ddots &  & \vdots \\ 
        \vdots & \vdots &  & \ddots & -{k} \\ 
        0 & 0 & \cdots & -{k} & {k}+{k}_\mathrm{g} 
    \end{bmatrix}_{{N_m}\times{N_m}},
\end{equation*}
represent the mass, damping, and the stiffness, matrices, respectively. Here, parameters $\{{m},{c},{k},{k}_\mathrm{g}\}$ represent the periodic mass, damping, interaction, and grounding stiffness properties of the baseline grounded monoatomic PnC. Parameters $\{{m}_\mathrm{def},{k}_\mathrm{g,def}\}$ represent the defect mass, and grounding stiffness at the defect location (i.e., the fluid--D-Psub interface). The linear ODE is first converted into a discrete state-space representation using the MATLAB command \texttt{c2d}, and subsequently integrated in time using an explicit Euler scheme at each time step.

The dynamical response of the defect mode can be additionally characterized through the undamped ($c=0$) velocity response of the defect mass $\dot{y}_1$ to harmonic forcing at the resonance frequency $\omega_\mathrm{def}$, which grows linearly in time as $\dot{y}_1 = \mathrm{A}_\mathrm{E} t \sin(\omega_\mathrm{def} t)$. The amplitude envelope $\mathrm{A}_\mathrm{E}$ quantifies this linear growth rate and serves as a measure of the intrinsic amplification of the D-Psub prior to coupling with the flow \cite{RamakrishnanJSV2025,RamakrishnanJFS2026}. 
Both the defect resonance frequency $\omega_\mathrm{def}$ and the amplitude envelope $\mathrm{A}_\mathrm{E}$ are governed by the mass and stiffness properties of the D-Psub. In this study, we vary the defect parameters $\{m_\mathrm{def}, k_\mathrm{g,def}\}$ to systematically vary these quantities and design phononic subsurfaces with prescribed spectral characteristics that are conducive to effective fluid--structure interaction.

It is important to note that the utility of the defect resonance \citep{RamakrishnanJSV2025} in our turbulent flow FSI configuration is significantly different from truncation resonances as reported in prior studies \citep{Hussein15,MichelisPoF2023}. Previous efforts designed PnCs to specifically leverage the out-of-phase surface interaction of the PnCs with a range of frequencies between the truncation resonance frequency and the upper edge of the phononic band gap. This was a sound phononic design strategy for effective FSI in scenarios involving relatively narrow-band flow coherences, e.g., Tollmien--Schlichting waves. However, given the fundamentally broadband and stochastic nature of a turbulent flow, the motivation for designing a defect resonance in PnCs is fundamentally different. Aligning a defect resonance with the desired turbulent flow scales allows us to realize a relatively narrow-band surface response at the fluid--D-Psub interface, even with a broadband fluid forcing applied on the channel wall. In this regard, the existence of a phononic band gap surrounding the prescribed defect resonance plays a critical role in filtering out the other non-resonant band gap frequencies in the flow, while allowing the D-Psub to manifest a dominant response close to the prescribed defect mode frequency.

\subsection{Turbulent channel flow simulation}

We perform direct numerical simulations (DNS) of the coupled turbulent channel flow over an array of phononic subsurfaces. The channel flow is driven by a constant mass flow rate at a friction Reynolds number $Re_\tau = u_{\tau,0} h/\nu \approx 186$, where $u_{\tau,0}$ is the friction velocity of the uncontrolled (rigid wall) case, $h$ is the channel half-height, and $\nu$ is the kinematic viscosity. 
The governing equations are given by the incompressible Navier--Stokes equations,
\begin{equation}
    \frac{\partial u_i}{\partial t}+u_j\frac{\partial u_i}{\partial x_j}
    =-\frac{1}{\rho}\frac{\partial p}{\partial x_i}
    +\nu \frac{\partial^2 u_i}{\partial x_j\partial x_j},\quad \frac{\partial u_i}{\partial x_i}=0
\end{equation}
where $x_1$, $x_2$, and $x_3$ denote the streamwise, wall-normal, and spanwise directions, respectively, $u_1$, $u_2$, and $u_3$ are the corresponding velocity components, $p$ is the pressure, and $\rho$ is the density. Periodic boundary conditions are applied in the streamwise and spanwise directions. At the top and bottom walls, no-slip conditions are imposed on the wall-parallel velocity components. The wall-normal velocity component $u_2$ at the top wall is given by a no-penetration condition, and at the bottom wall is prescribed by the weakly coupled scheme described in Section \ref{sec:weakly}.

The turbulent channel flow is solved using a staggered-grid, second-order finite-difference discretization in space, with a fractional-step method for pressure-velocity coupling, and a third-order Runge--Kutta time integration scheme. The computational domain has dimensions $(L_1/h, L_3/h) = (4\pi, 2\pi)$, where $L_1$ and $L_3$ are the domain size in the streamwise and spanwise directions, respectively, with uniform grid spacing in the homogeneous directions ($\Delta x_1^+ \approx 10$, $\Delta x_3^+ \approx 5$) and a hyperbolically stretched wall-normal grid ($\Delta x^+_{2,\min} \approx 0.16$ near the wall and $\Delta x^+_{2,\max} \approx 7.3$ at the center line). The domain size is big enough to capture the effect of the blowing-and-suction boundary condition on the turbulent statistics~\citep{lin_24}. Note that superscript $(\cdot)^+$  denotes quantities normalized by $u_{\tau,0}$, $\rho$, and $\nu$. A fixed time step of $\Delta t^+ \approx 0.01$ is used, and the total integration time is $T^+ \approx 8000$ to ensure convergence of time-averaged statistics. Code validation is achieved in prior studies of uncontrolled turbulent channel flow and controlled channel flow as well~\citep{bae_2018, bae_2019, Lozano_Duran_2016,lin_24}.

In order to solve for pressure, the pressure field $p$ is decomposed into mean pressure gradient $\bar{p}$ and pressure fluctuation $p'$. The imposed mean pressure gradient $\bar{p}$ is adjusted to maintain the constant mass flow rate, while the pressure fluctuation is computed via the pressure Poisson equation. In this study, we adopt the simplified form of the pressure Poisson equation used in the previous unsteady wall-transpiration case by \citet{Toedtli_Leonard_McKeon_2025}, where pressure fluctuation is given by
\begin{equation}
    \frac{1}{\rho}\frac{\partial^2 p'}{\partial x_i \partial x_i}=-\frac{\partial u_i}{\partial x_j}\frac{\partial u_j}{\partial x_i},
    \label{eq:PPE equation}
\end{equation}
with pressure boundary condition given by
\begin{equation}
     \frac{1}{\rho} \left.\frac{\partial p'}{\partial x_2}\right|_{x_2=0,2h}= \nu \frac{\partial^2 u_2}{\partial x_i \partial x_i}-\frac{\partial u_2}{\partial t}.
\end{equation}

\begin{figure*}[t]
  \centering
  \includegraphics[width=1\textwidth]{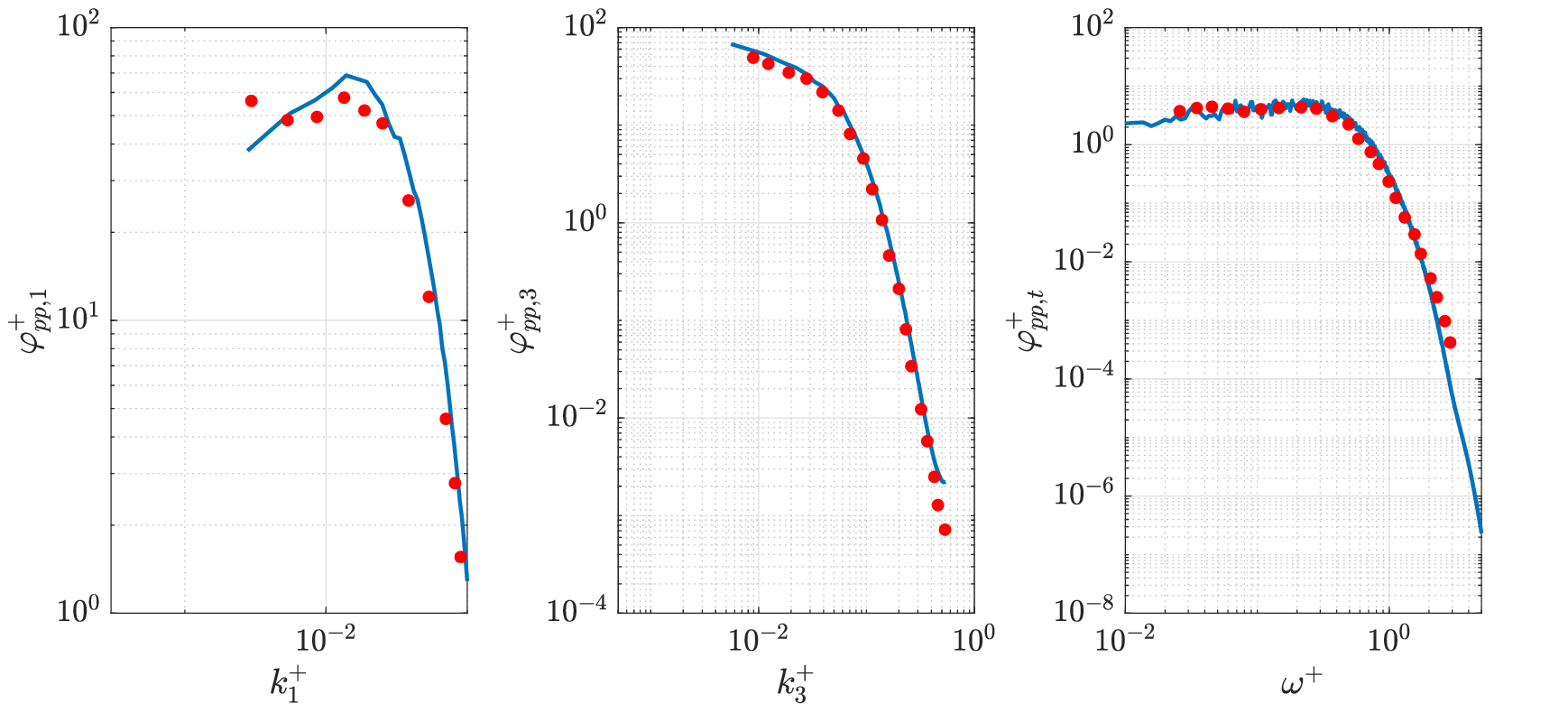}
  \caption{(a) Streamwise, (b) spanwise, and (c) temporal wall-pressure power spectral density of the uncontrolled turbulent channel flow at $Re_\tau = 186$ from the current DNS (blue line) and \citet{Choi_and_Moin_1990} (red circle).}
  \label{fig:wall_pressure_spectra}
\end{figure*}

Fig.~\ref{fig:wall_pressure_spectra} presents the wall-pressure power spectrum with no penetration boundary condition at the bottom wall. The wall-pressure fluctuation spectrum is defined in the three-dimensional spectral space as
\begin{equation}
    \varphi_{pp}(k_1,k_3,\omega) = 
    \frac{\hat{p'}(k_1, x_2 = 0, k_3, \omega)\,\hat{p'}^\dagger(k_1, x_2 = 0, k_3, \omega)}
    {\Delta k_1\,\Delta k_3\,\Delta \omega},
\end{equation}
where $\hat{p'}(k_1, x_2=0, k_3, \omega)$ denotes the Fourier transform of the wall-pressure fluctuation in the streamwise, spanwise and temporal directions, and $(\cdot)^\dagger$ represents the complex conjugate. Here, $k_1$ and $k_3$ denote the streamwise and spanwise wavenumbers, respectively, and $\omega$ is the temporal angular frequency. 
The corresponding one-dimensional streamwise, spanwise, and temporal spectra, $\varphi_{pp,1}$, $\varphi_{pp,3}$ and $\varphi_{pp,t}$, are obtained by integrating $\varphi_{pp}(k_1,k_3,\omega)$ over the remaining two dimensions. The resulting spectra in streamwise and spanwise wavenumbers, as well as temporal frequency, exhibit broadband distributions that are consistent with previous DNS studies \citep{Choi_and_Moin_1990,Yang_Yang_2022}. 

\subsection{Weakly-coupled D-Psub boundary condition} 
\label{sec:weakly}

The interaction between the phononic subsurfaces and the turbulent channel flow is updated sequentially at each time step, where the wall-pressure is computed to provide the forcing to the D-Psub, and the D-Psub displacement is calculated to provide the boundary condition for the channel flow.

The bottom wall is sectioned into $N_{\mathrm{DPsub}}$ control panels in the streamwise direction such that each D-Psub panel spans the full spanwise length of the domain as shown in Fig.~\ref{fig:schematic}. The mass of the D-Psub is aligned with the wall-normal direction, such that the force $F_j$ applied to the $j$-th D-Psub is the wall-normal force on the interface and is given by the spatial average of the pressure on each panel, 
\begin{equation}
    F_j(t) = \int_{(j-1)\lambda_1}^{j\lambda_1}\int^{L_3}_0p(x_1,\;x_2=0,\;x_3,\;t) \mathrm{d}x_3\mathrm{d}x_1,\quad 1\le j\le N_{\mathrm{DPsub}},
\end{equation}
where $\lambda_1$ is the streamwise length of the D-Psub control panel.

Once the net normal load $F_j$ is computed, at the following time step, the computed load is applied as a temporal forcing to the D-Psub model (Eq.~\eqref{eq:RPM_equation}) for each D-Psub panel. The structural response is recorded solely in terms of the velocity of the interface (defect) mass, $V_{m,j}(t)$, without explicitly computing the wall displacement field. This interface velocity is then mapped back into the fluid domain as a spatially localized, streamwise Gaussian profile that enforces a zero-net-flux transpiration boundary condition on each D-Psub panel defined as
\begin{equation}
    u_2(x_1, x_2=0, x_3,t)=V_{m,j}(t)\cos\left[\frac{3\pi}{\lambda_1}\left(x_1-\left(j-\frac{1}{2}\right)\lambda_1\right)\right]\exp\left[{-\frac{\left(x_1-\left(j-\frac{1}{2}\right)\lambda_1\right)^2}{2\sigma^2}}\right],\, (j-1)\lambda_1 \le x_1 \le j\lambda_1,
\end{equation}
where the width $\sigma$ is determined numerically to satisfy the net-zero flux condition for each panel.

Using wall velocity to approximate wall roughness has been successfully employed in studies of riblets, where the influence of surface geometry can be represented through virtual origins \citep{Ibrahim_21,GOMEZDESEGURA2020108675}. Extending to unsteady motion, we approximate the wall-normal deformation of the phononic subsurface through its corresponding wall-normal velocity, assuming that the wall displacement remains sufficiently small compared with the characteristic flow length scales \citep{Min_2006,Toedtli_Leonard_McKeon_2025}. Under this small-displacement assumption, the leading-order term of wall motion on the flow can be represented by a transpiration boundary condition imposed on a no-slip wall. However, when the deformation amplitude becomes large, the equivalence between wall motion and transpiration may break down. For example, \citet{HEPFFNER_FUKAGATA_2009} showed that traveling-wave wall deformation and blowing-suction actuation can produce opposite pumping directions when the deformation amplitude is sufficiently large. To ensure that the present approximation remains valid, we therefore constrain the maximum displacement of the D-Psub models to remain within the buffer layer, i.e., $x_2^+ \le 25$.

\section{Phononic subsurface dynamics parameter selection}\label{sec:RPM_selection}

To design the phononic subsurface for effective interaction with turbulent flow, it is necessary to identify the characteristic spatial and temporal scales of the flow. However, owing to the broadband nature of turbulent spectra across both spatial and temporal scales, it is difficult to isolate a single target scale for phononic subsurface design. Therefore, to determine suitable design parameters, we rely on a prior analysis of harmonic blowing and suction in turbulent channel flow~\citep{lin_24}, the parameter space was systematically explored with the objective of maximizing drag reduction. In the present work, we adopt the optimal wavelength--frequency combinations identified from these contour maps to guide the design of the phononic subsurface.

A detailed analysis of these optimal spectral scales has been reported in our prior work~\citet{lin_24}. In this study, a blowing and suction boundary condition (i.e., no wall deformation but non-zero wall velocity) was applied at the channel walls, with a prescribed standing wave-type wall-normal wall velocity:
\begin{equation}
    u_2(x_1, x_2=0, x_3, t)=A\cos\left(\frac{2\pi x_1}{\lambda}\right)\cos\left(\omega t\right),
\end{equation}
and perform comprehensive parameter sweeps of the wall velocity amplitude, $A$, the streamwise wavelength, $\lambda$, and the temporal oscillation frequency, $\omega$, to identify optimal wall velocity parameters conducive to turbulent drag reduction. 

The prescribed standing-wave study identifies the optimal parameters for drag reduction as $(A^+_\mathrm{opt}, \lambda^+_\mathrm{opt}, \omega^+_\mathrm{opt}) = (0.7,\,206,\,0.1)$, corresponding to a maximum drag reduction of approximately $2.93\%$ relative to the baseline rigid-wall turbulent channel flow. 
Based on these target values, the streamwise length of the phononic subsurface is fixed at $\lambda_1^+ = 206$, which corresponds to $N_{\mathrm{DPsub}} = 11$ in the present computational domain.

To ensure a sufficiently wide band gap around the target frequency $\omega^+ \approx 0.1$, the defect material properties are selected to span the dynamically relevant range. A parametric study is conducted by varying the defect mass and stiffness, $(m_{\mathrm{def}},\; k_{\mathrm{g,def}})$, within $m^+_{\mathrm{def}} = {m_{\mathrm{def}} u_{\tau,0}^3}/({\rho \nu^3}) \in (4.5,\,12.87)\times 10^6$ and $k^+_{\mathrm{g,def}} ={k_\mathrm{g,def}}/({\rho \nu u_{\tau,0}})\in (0.5,\,19.8)\times 10^{4}$. 
In total, 34 D-Psub configurations are considered under the assumption of small wall displacement. The remaining parameters are held fixed as follows: $m^+=6.4\times 10^6$, $k^+=5.6\times 10^{4}$, $k_g^+=1.4\times 10^{6}$, and $c^+={c u_{\tau,0}}/({\rho \nu^2})=3.0\times10^3$. The complete set of parameters, along with the corresponding flow and structural response metrics, is summarized in Appendix~\ref{sec:appendix_RPM}.

By performing a parametric sweep over the defect mass and grounding stiffness, a range of designed dynamics can be obtained in terms of the amplitude envelope $A_E$ and the defect frequency $\omega_{\mathrm{def}}$. These two quantities represent the uncoupled dynamics of the phononic subsurfaces, as defined in Sec.~\ref{sec:RPM}, and are designed to interact with turbulent flow at the relevant temporal scales identified from the prior prescribed blowing and suction study. Both quantities are summarized in Appendix~\ref{sec:appendix_RPM} and illustrated as contour lines in Fig.~\ref{fig:A_omega_map}. Note that $A_E^+={A_E \nu}/{u_{\tau,0}^3}$ and $A_m^+={A_m \nu}/{u_{\tau,0}^3}$.

The amplitude envelope exhibits a strong dependence on the defect mass $m_{\mathrm{def}}$ and only a weak dependence on the defect stiffness $k_{g,\mathrm{def}}$. This behavior suggests that, within the considered parameter range, the amplification mechanism of the undamped phononic subsurface is primarily governed by the inertia of the defect mass. In contrast, the defect frequency increases with increasing stiffness $k_{g,\mathrm{def}}$ and decreasing mass $m_{\mathrm{def}}$, consistent with the scaling of a one-degree-of-freedom mass--spring system, $\omega \sim \sqrt{k/m}$.

However, because the phononic subsurfaces are coupled with turbulent flow, these uncoupled design parameters alone are insufficient to fully characterize the system behavior. Therefore, in the next section, we examine the coupled dynamics as functions of defect mass and stiffness.

\section{Coupled D-Psub Dynamics}\label{sec: RPM dynamics}
In this section, we shift the focus from the designed (uncoupled) dynamics to the coupled dynamics in order to examine how the D-Psubs respond to turbulent flow. The D-Psub's response can be characterized from three fundamental perspectives: the coupled amplitude, the dominant frequency, and the phase difference between adjacent panels. Together, these quantities provide a comprehensive description of the D-Psub motion under the influence of turbulent channel flow. 

\subsection{Coupled amplitude and frequency}\label{sec: A_m and omega}
\begin{figure*}[t]
    \centering

    \begin{minipage}[t]{0.48\textwidth}
        \begin{overpic}[width=\textwidth]{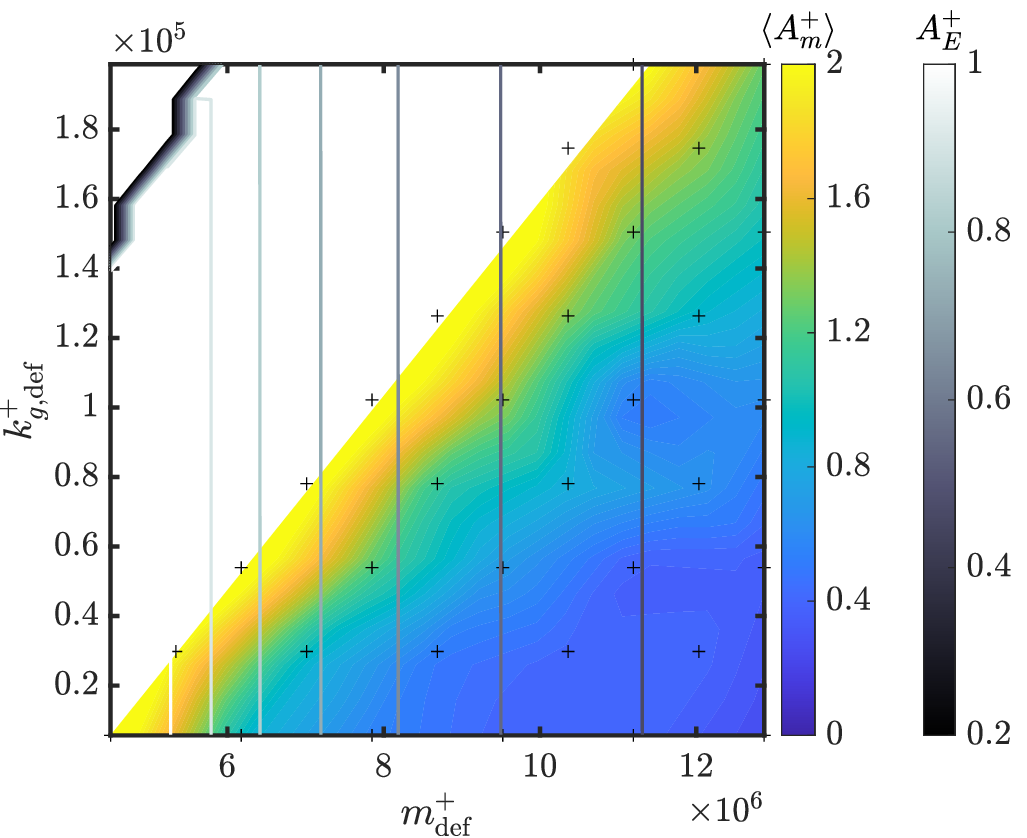}
            \put(16,81){\textbf{(a)}}  
        \end{overpic}
    \end{minipage}%
    \hfill
    \begin{minipage}[t]{0.48\textwidth}
        \hfill
        \begin{overpic}[width=\textwidth]{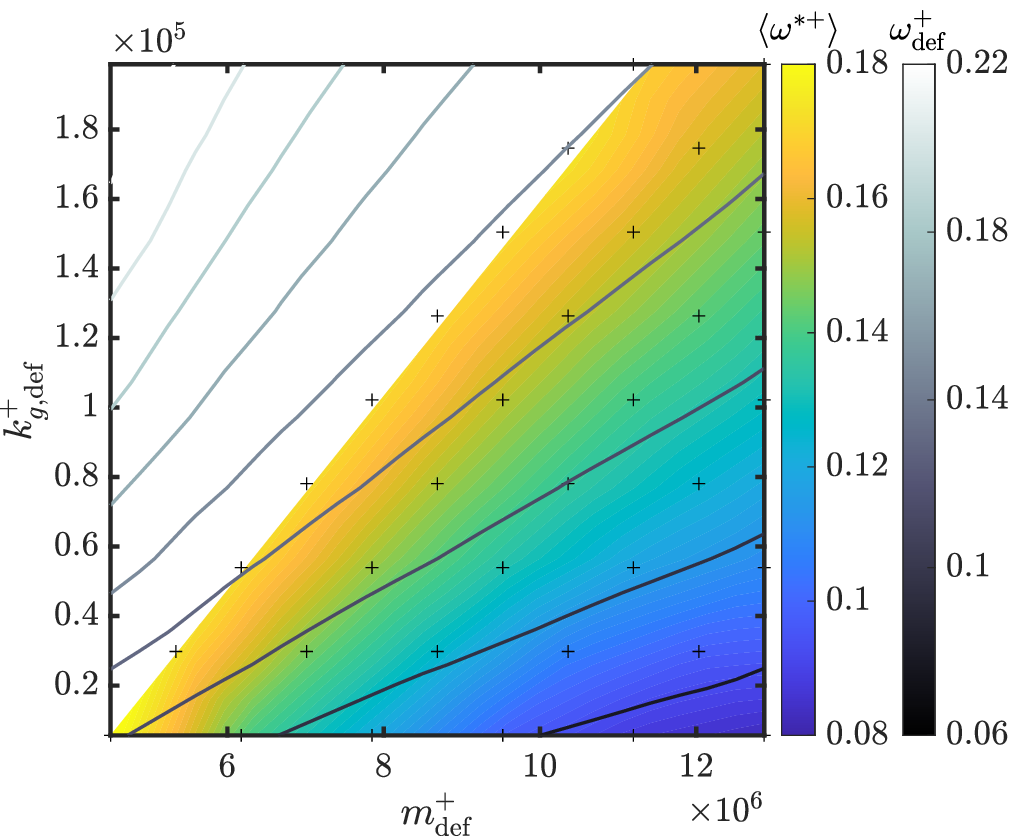}
            \put(16,81){\textbf{(b)}}  
        \end{overpic}
    \end{minipage}

    \caption{(a) Contour maps of the panel-averaged velocity amplitude $\langle A_m \rangle$ (color) and amplitude envelope $A_E$ (line) and (b) panel-averaged peak frequency $\langle \omega^{*} \rangle$ (color) and defect frequency $\omega_{\mathrm{def}}$ (line) as functions of the D-Psub parameter pair $(m_{\mathrm{def}}, k_{\mathrm{g,def}})$. The symbols ($+$) denote the parameter locations corresponding to the numerical experiments.}

    \label{fig:A_omega_map}
\end{figure*}
We first evaluate the coupled amplitude and dominant oscillation frequency in Fig.~\ref{fig:A_omega_map}. The coupled amplitude $A_m$ is defined as the peak-to-peak amplitude of the center mass oscillating velocity $V_{m,j}(t)$ for each panel. The dominant frequency $\omega^{*}$ is determined from the peak of the power spectrum of $V_{m,j}$. These two quantities are consistent across the control panels once transient effects are excluded, indicating a statistically homogeneous response. Therefore, in the following analysis, we focus on panel-averaged metrics, denoted by angle brackets, $\langle q \rangle = (1/N_{DPsub})\sum_{j=1}^{N_{\mathrm{DPsub}}} q_j$.

The contour map of the panel-averaged velocity amplitude $\langle A_m \rangle$ in Fig.~\ref{fig:A_omega_map}(a) reveals a clear monotonic trend: the amplitude increases as the defect mass $m_{\mathrm{def}}$ decreases and as the defect stiffness $k_{\mathrm{g,def}}$ increases. Superimposing the contour lines of the amplitude envelope, $A_E$, highlights a notable discrepancy between the two measures. Within the explored parameter space, $A_E$ remains largely insensitive to variations in defect stiffness, whereas $\langle A_m \rangle$ shows a pronounced dependence. This mismatch indicates that the amplitude envelope cannot serve as an a priori predictor of the equilibrium oscillation amplitude in the fully coupled fluid--metamaterial interaction. Instead, the realized velocity amplitude emerges from the nonlinear interaction between structural dynamics and turbulent forcing.

The dominant frequency map in Fig.~\ref{fig:A_omega_map}(b) exhibits a similar parametric dependence. The panel-averaged peak frequency $\langle \omega^{*} \rangle$ increases with decreasing $m_{\mathrm{def}}$ and increasing $k_{\mathrm{g,def}}$, consistent with the expected scaling of the structural natural frequency. The dominant frequency generally follows the designed defect frequency $\omega_{\mathrm{def}}$. However, a non-negligible discrepancy is observed between the two, with differences of up to approximately $30\%$. Such deviations are sufficiently large to produce significant variations in the resulting drag response. The underlying mechanism and spectrum analysis are investigated in Section~\ref{sec: freq shift}
\subsection{Phase relation between phononic subsurfaces}
\begin{figure}
    \begin{minipage}[t]{0.29\textwidth}
        \begin{overpic}[width=\textwidth]{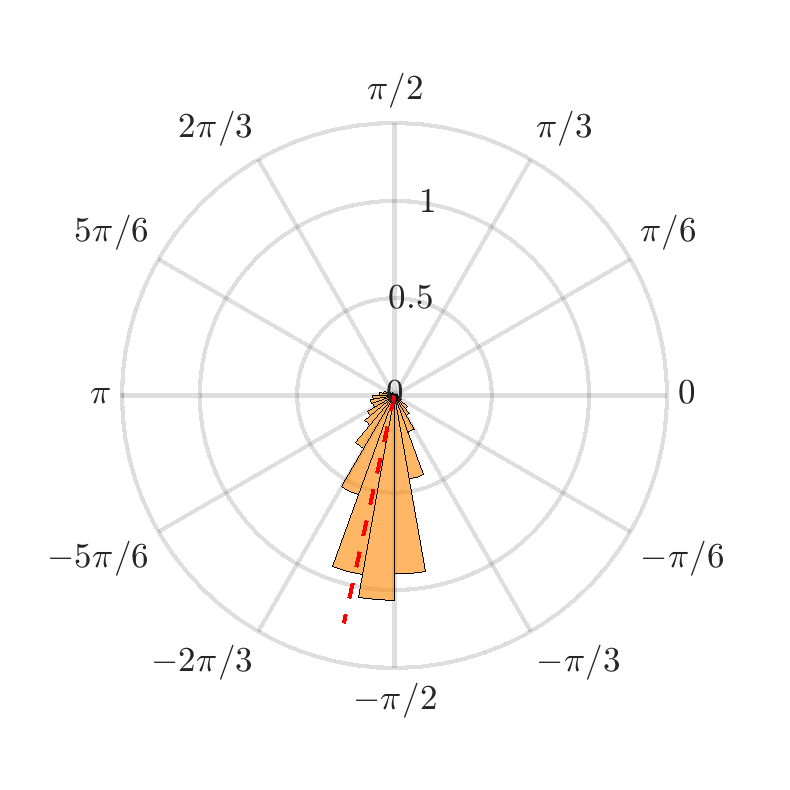}
            \put(16,92){\textbf{(a)}}  
        \end{overpic}
    \end{minipage}%
    \hfill
    \begin{minipage}[t]{0.34\textwidth}
        \hfill
        \begin{overpic}[width=\textwidth]{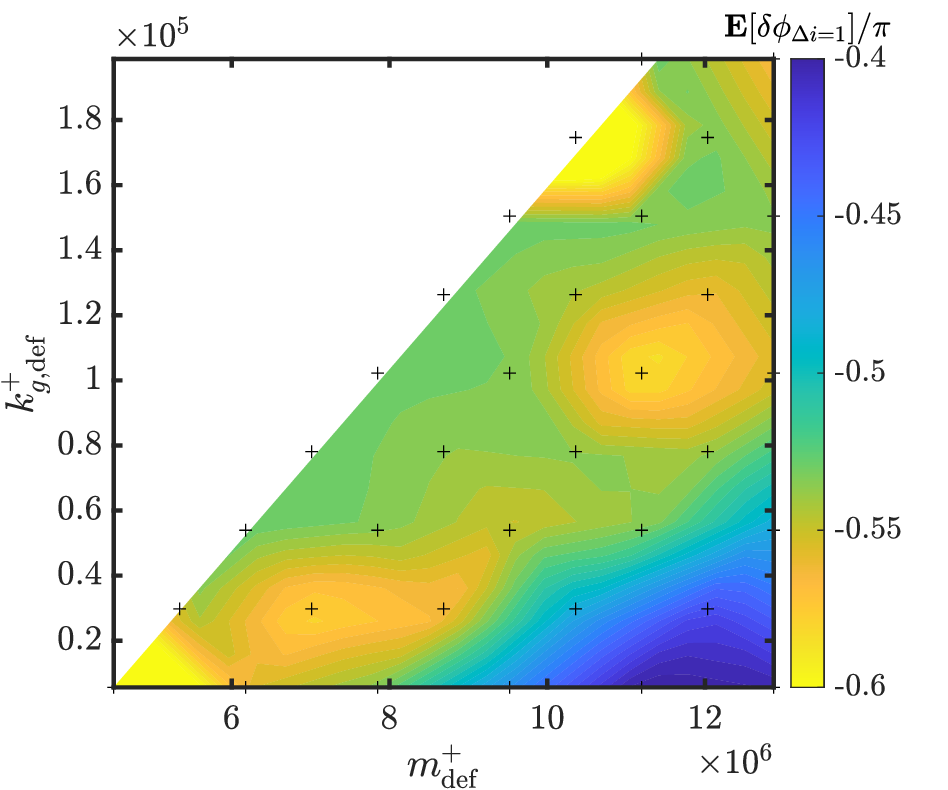}
            \put(23,80){\textbf{(b)}}  
        \end{overpic}
    \end{minipage}
    \hfill
    \begin{minipage}[t]{0.34\textwidth}
        \hfill
        \begin{overpic}[width=\textwidth]{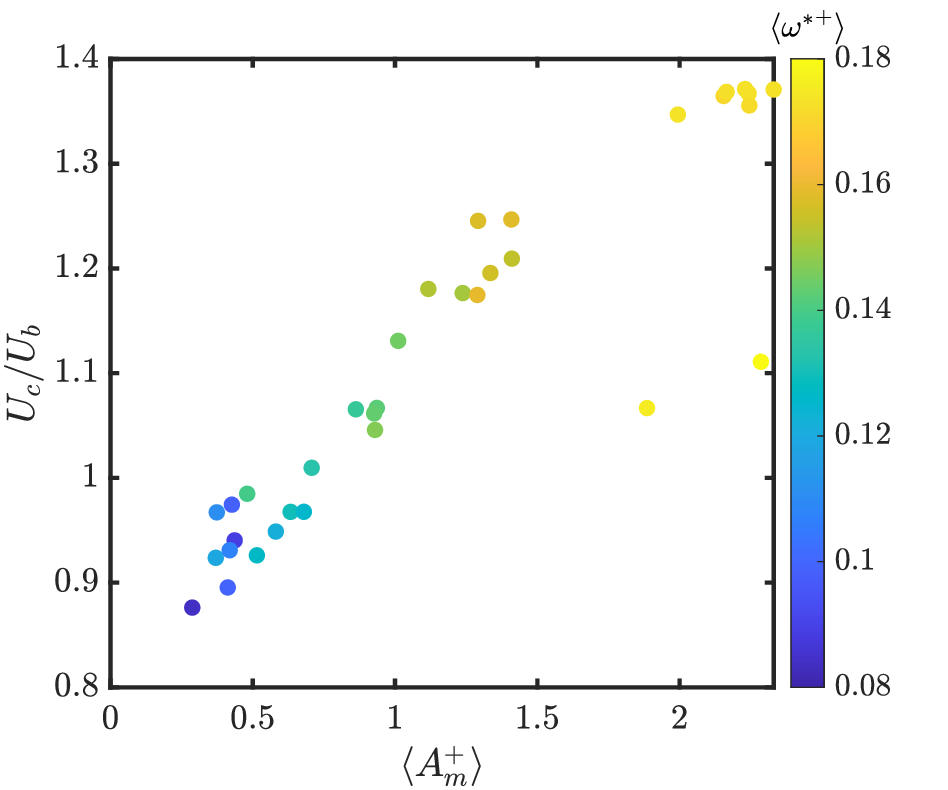}
            \put(16,80){\textbf{(c)}}  
        \end{overpic}
    \end{minipage}
    \caption{(a) Probability density function $\rho(\delta \phi_{\Delta i = 1}(t))$ of the phase difference between adjacent panels for case~10. The red dashed line denotes the expected value of the phase difference. 
    (b) Contour map of the expected value of phase shift between adjacent panels as a function of the D-Psub parameter pair $(m_{\mathrm{def}},\, k_{\mathrm{g,def}})$. The symbols ($+$) denote the parameter locations corresponding to the numerical experiments. 
    (c) Scatter plot of the convection velocity $U_c$ scaled by bulk velocity $U_b$ versus the equilibrium velocity amplitude $\langle A_m \rangle$, colored by the dominant frequency ${\omega^*}$.}

    \label{fig:Vm_phase}
\end{figure}

Furthermore, the phase relationship among the D-Psubs is examined to better characterize the collective dynamics of the coupled system. Because each D-Psub responds locally and independently to the wall-normal pressure loading above it, phase differences among adjacent panels naturally arise. If the perturbations induced by the D-Psubs are convected downstream at a finite convection velocity, these disturbances may influence downstream subsurfaces with a systematic phase offset. Understanding these phase relations is essential for characterizing the unconstrained collective motion of the D-Psubs, beyond single-panel dynamics.

To quantify the phase relationship among the unsteady motions of multiple D-Psubs, the Hilbert transform is employed to extract the instantaneous phase from the subsurface velocity signals. For a narrow-band time-series signal $\mathrm{x}(t)$, the Hilbert transform constructs the corresponding analytic signal $\mathcal{Q}(t)$ as \citep{Matsuki2023,Marple99}.
\begin{equation}
    \begin{split}
        &\mathcal{Q}(t)={q}(t)+\mathrm{i}H[{q}(t)],\\
        &H[{q}(t)]=\pi^{-1}\mathrm{P.V.}\int^\infty_{-\infty}\frac{{q}(\tau)}{t-\tau}\,\mathrm{d}\tau,
    \end{split}
\end{equation}
where $\mathrm{P.V.}$ is the Cauchy principal value and function $H[\cdot]$ denotes the Hilbert transform. The instantaneous phase of the signal is then obtained from the argument of the analytic signal,
\begin{equation}
    \phi_{{q}}(t)=\angle (\mathcal{Q}(t)).
    \label{eq:phase_hilbert}
\end{equation}

The instantaneous phase of the centerline velocity $V_{m,j}$ for panel index $j$ is extracted using Eq.~\eqref{eq:phase_hilbert}, with the time signal ${q}(t)$ replaced by $V_{m,j}(t)$. Owing to the streamwise periodicity of the computational domain and the uniform spacing of the D-Psub panels, the phase relationship can be conveniently characterized in terms of the relative panel index. Specifically, the phase difference between two panels separated by an index offset $\Delta i$ is defined as
\begin{equation}
    \begin{split}
        \delta \phi_{\Delta i} (t) = \left\{\phi_{V_{m,j+\Delta i}}(t)-\phi_{V_{m,j}}(t): 1\le j \le N_{\mathrm{DPsub}} \right\},
    \end{split}
\end{equation}
which represents the ensemble of phase differences corresponding to all panel pairs separated by the same relative distance $\Delta i$.

After discarding the initial transient stage, the subsurface velocity signals are analyzed over statistically stationary time intervals. The probability distribution function, $\rho(\delta \phi_{\Delta i} (t) )$, is then computed to characterize the statistical phase relationships among phononic subsurfaces, providing insight to the general behavior of the control panels reacting passively to a broad-band and chaotic turbulent flow.

Fig.~\ref{fig:Vm_phase}(a) presents the polar histogram of probability distribution function, $\rho(\delta \phi_{\Delta i =1} (t) )$, for a representative configuration (Case~10). We observe that the downstream panel lags the upstream panel by approximately $\pi/2$, and the relatively narrow distribution indicates a strong phase correlation between neighboring panels over the control horizon. As the separation increases downstream, the expected value of  phase difference approximately scales with the panel index separation, suggesting a coherent convective influence. At the same time, the progressively broader distributions observed for larger $\Delta i$ indicate increasing variability and more chaotic relative motion among the panels that are farther separated in space. 

Fig.~\ref{fig:Vm_phase}(b) presents the contour map of the expected phase shift as a function of the material parameter pair $(m_{\mathrm{def}}, k_{\mathrm{g,def}})$. The phase shift is predominantly negative across the parameter space, indicating a phase lag between downstream adjacent panels. This behavior is consistent with a convective transport mechanism, whereby fluid responses propagate downstream following the mean flow. The magnitude of the phase lag therefore reflects an effective convection speed associated with the coupled dynamics.

To establish the prediction model for the phase difference across adjacent panels, we adopt a convection velocity model, in which we can assume that the  information is convected downstream at a constant speed $U_c$, the expected value of phase difference between adjacent panels can be expressed as
\begin{equation}
\mathbb{E}[\delta \phi_{\Delta i=1}] = - \frac{\lambda_1\omega^*}{U_c} \in [-\pi,\pi].
\label{eq:E_phase_eq}
\end{equation}
We then compute the convection velocity $U_c$ and present the results in Fig.~\ref{fig:Vm_phase}(c). We see that $U_c/U_b\approx 1$, where $U_b^+ \approx 15.88$ is the bulk velocity. This indicates that under weak blowing and suction, the bulk velocity serves as the dominant convection velocity in the system. We also observe that for $\langle A_m^+ \rangle < 1.5$, the convection velocity increases approximately linearly with $\langle A_m \rangle$. This regime corresponds to moderate structural response and aligns with the drag-reduction region discussed in the following section.

For larger amplitudes and higher dominant frequencies, the relationship deviates from linearity, and demonstrates a bifurcation-like behavior. In this regime, multiple convection characteristics emerge for comparable amplitudes, indicating that the convection dynamics are not uniquely determined by the equilibrium oscillation amplitude. Instead, they reflect the influence of nonlinear fluid--structure coupling and interactions across multiple dynamical scales. Nevertheless, the strong linear correlation observed in the moderate-amplitude regime suggests that the coupled amplitude still serves as a reliable predictor of the phase difference between adjacent panels.

\subsection{Dominant frequency shift}\label{sec: freq shift}
In Section~\ref{sec: A_m and omega}, we showed that the dominant frequency of the phononic subsurface, $\omega^{*}$, in the coupled system exhibits a discrepancy relative to the designed defect frequency $\omega_{\mathrm{def}}$. In this section, we examine in detail the mechanisms responsible for this difference. To quantify the deviation of the coupled response from the designed structural frequency, we define the frequency shift
\begin{equation}
\Delta \omega = {\omega^{*}} - \omega_{\mathrm{def}} ,
\end{equation}
where ${\omega^{*}}$ denotes the dominant coupled oscillation frequency and $\omega_{\mathrm{def}}$ is the nominal defect frequency.
\begin{figure*}[t]
    \centering

    \begin{minipage}[t]{0.48\textwidth}
        \hfill
        \begin{overpic}[width=\textwidth]{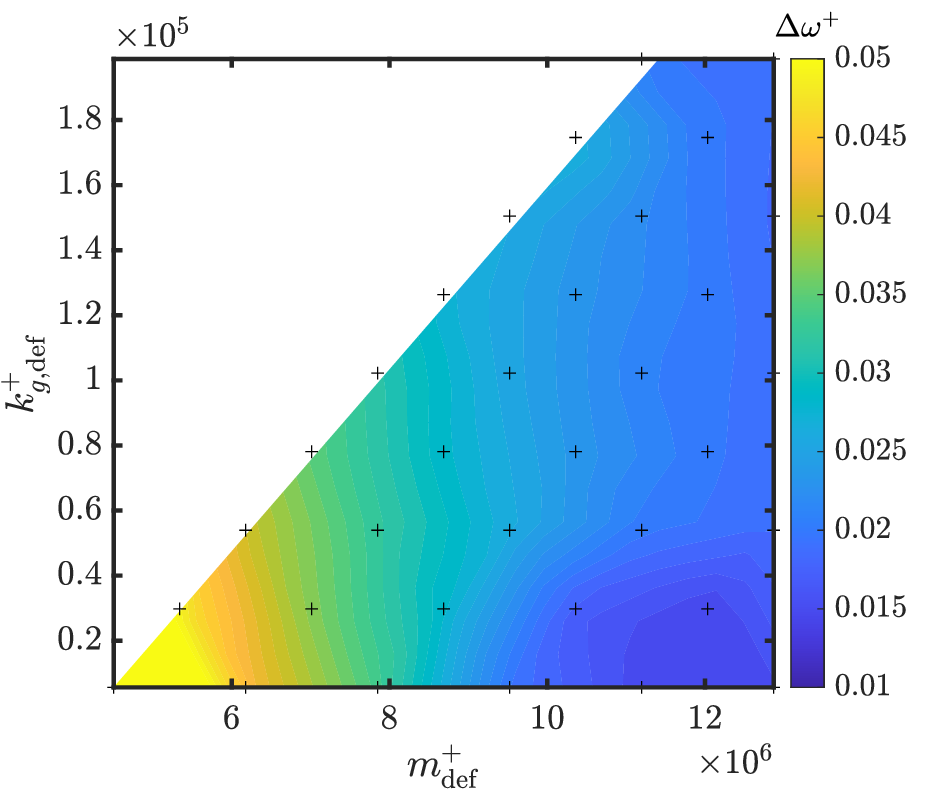}
            \put(22,80){\large\textbf{(a)}}  
        \end{overpic}
    \end{minipage}
    \hfill
    \begin{minipage}[t]{0.48\textwidth}
        \hfill
        \begin{overpic}[width=\textwidth]{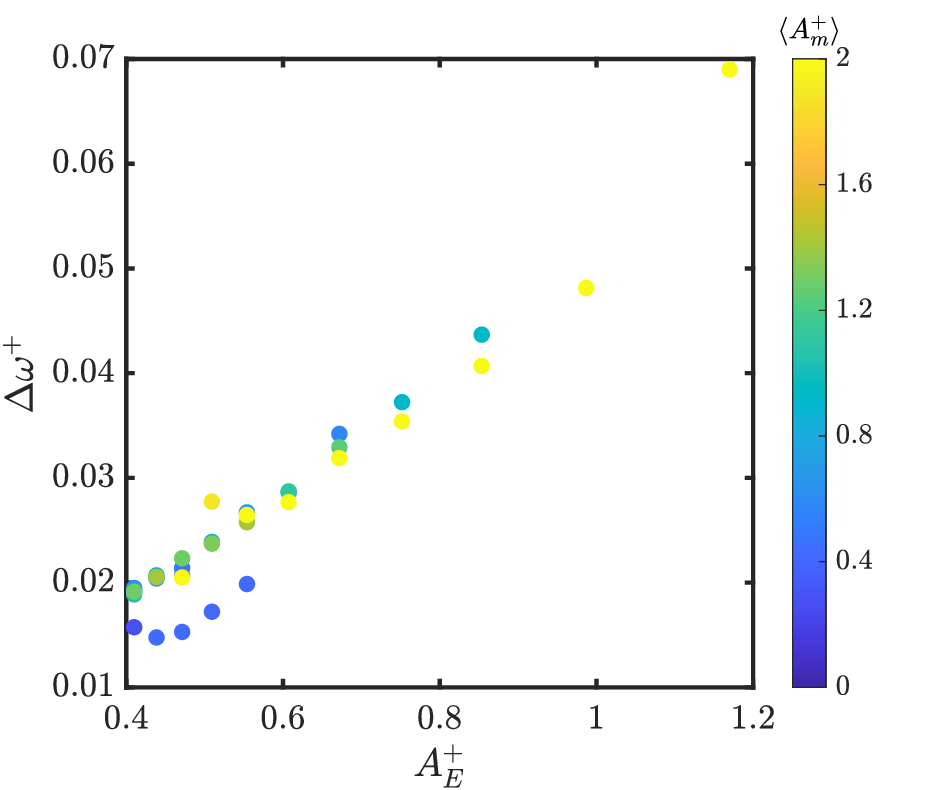}
            \put(16,80){\large\textbf{(b)}}  
        \end{overpic}
    \end{minipage}

    \caption{(a) Contour map of the frequency shift $\Delta \omega$ as a function of the D-Psub parameter pair $(m_{\mathrm{def}}, k_{\mathrm{g,def}})$. The symbols ($+$) denote the parameter locations corresponding to the numerical experiments. (b) Scatter plot of $\Delta \omega$ versus the amplitude envelope $A_E$, colored by the panel-averaged velocity amplitude $\langle A_m \rangle$.}

    \label{fig:DW_map}
\end{figure*}

The contour map of $\Delta \omega$ in Fig.~\ref{fig:DW_map}(a) illustrates its dependence on the material parameter pair $(m_{\mathrm{def}}, k_{\mathrm{g,def}})$. For configurations with relatively large defect mass and small defect stiffness, the frequency shift remains weak. In these cases, the structural response amplitude is small, and the limited wall-normal motion induces only a minor frequency shift in the coupled system. As the defect mass decreases, $\Delta \omega$ increases systematically, whereas its sensitivity to defect stiffness is comparatively weak within the explored parameter range.

To further interpret this behavior, Fig.~\ref{fig:DW_map}(b) presents $\Delta \omega$ as a function of the amplitude envelope $A_E$. A clear linear relationship emerges between the frequency shift and $A_E$. In contrast, the scatter color representing the panel-averaged velocity amplitude $\langle A_m \rangle$ does not reveal a strong correlation with $\Delta \omega$. This distinction indicates that the frequency shift is not directly governed by the realized oscillation amplitude, but rather by the linear amplification characteristics associated with the defect frequency.

\begin{figure}[t]
    \begin{minipage}[t]{0.48\textwidth}
        \hfill
        \begin{overpic}[width=\textwidth]{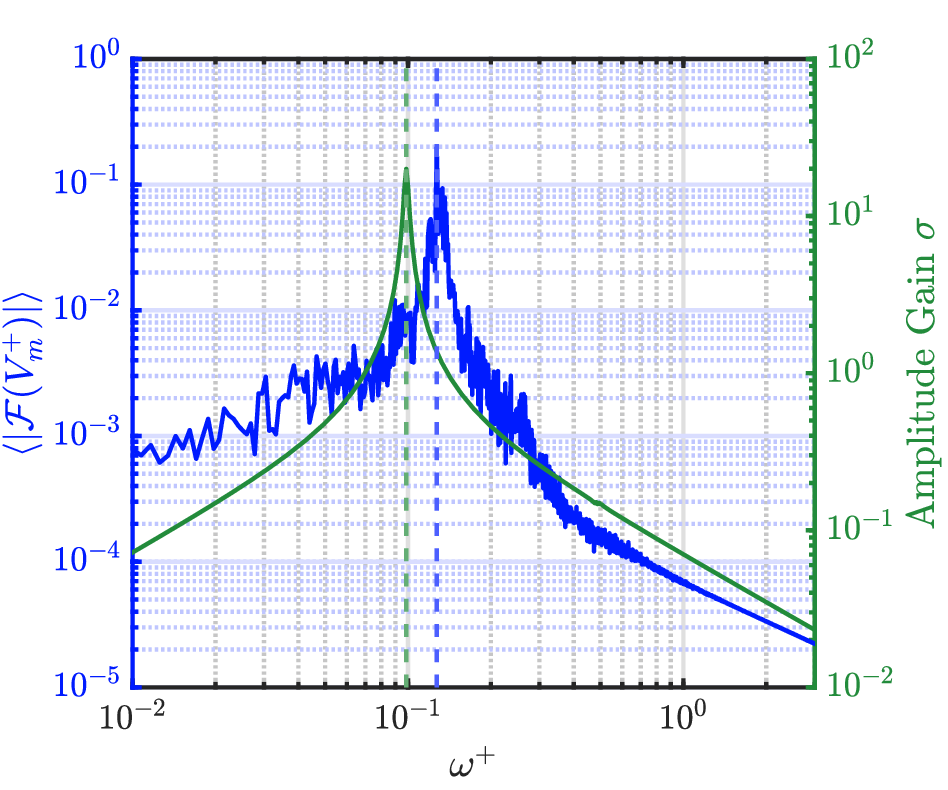}
            \put(16,79){\large\textbf{(a)}}  
        \end{overpic}
    \end{minipage}
    \hfill
    \begin{minipage}[t]{0.48\textwidth}
        \hfill
        \begin{overpic}[width=\textwidth]{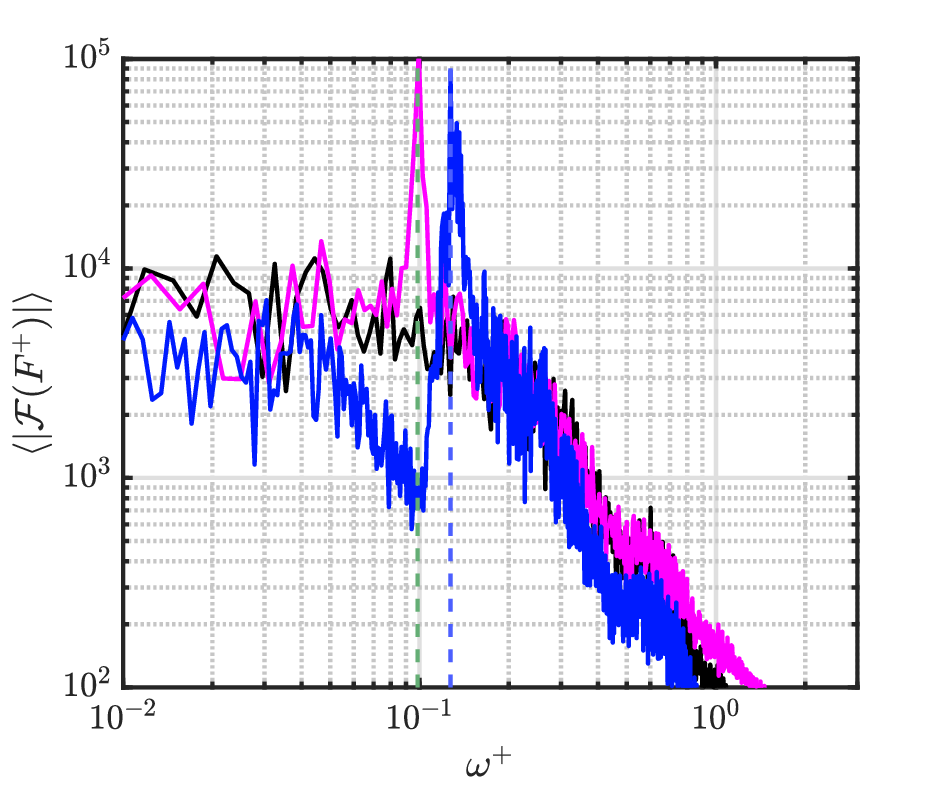}
            \put(16,79){\large\textbf{(b)}}  
        \end{overpic}
    \end{minipage}
    \caption{Spectra of the (a) panel-averaged peak velocity magnitude for case~10 (blue) and maximum gain (green)  and (b) wall-normal force for the controlled case~10 (blue), the uncontrolled case (black), and the case with a prescribed Gaussian forcing described in Eq.~\eqref{eq:pres_gauss} (magenta). Vertical dash lines indicate peak frequency corresponding to $\omega_{\mathrm{def}}$ (green) and $\omega^*$ (blue).}

    \label{fig:spectrum_39}
\end{figure}
These differences are also evident in the frequency spectrum of the centerline velocity $V_{m,j}$. For brevity, we present only a representative case (case~10), although the observed dynamics are consistent across different cases. Fig.~\ref{fig:spectrum_39} shows the spectrum of the coupled system, including the spectra of the panel-averaged surface velocity $\langle |\mathcal{F}(V_m)| \rangle$ and the wall-normal force $\langle |\mathcal{F}(F)| \rangle$ acting on the D-Psub, where $\mathcal{F}(\cdot)$ denotes the Fourier transform operator in time. Note that $V_m^+={V_m}/{u_{\tau,0}}$ and $F^+={F}/({\rho \nu^2})$.

In Fig.~\ref{fig:spectrum_39}(a), the spectrum of the panel-averaged centerline velocity magnitude exhibits a pronounced concentration of energy around the designed defect frequency of the D-Psub. This indicates that the subsurface structure selectively amplifies flow-induced motions near its targeted resonant band. Notably, the peak of the velocity spectrum is observed to shift slightly toward higher frequencies relative to the nominal defect frequency. Such a shift suggests that the effective resonance of the coupled fluid--structure system is modified by interaction with the turbulent flow.

Fig.~\ref{fig:spectrum_39}(b) compares the wall-normal force spectrum of the controlled flow with that of the uncontrolled reference case. The controlled spectrum reveals synchronization with the dominant frequency observed in the velocity response, which is expected, as the wall-normal pressure loading is directly coupled to the wall motion and therefore exhibits a similar spectral content. At the originally designed defect frequency, the wall-normal force spectrum exhibits a reduction relative to the uncontrolled case, consistent with sink-like behavior, indicating a local extraction of energy from the flow. This reduction is accompanied by a corresponding amplification at the newly established coupled peak frequency, suggesting a transfer of energy from the original defect band to the dominant coupled mode. 

To isolate the mechanism leading to the observed frequency shift, we impose a prescribed Gaussian blowing and suction profile,
\begin{equation}
    u_2(x_1, x_2=0, x_3,t)=A\cos(\omega t+j\delta\phi)\cos\left[\frac{3\pi}{\lambda_1}\left(x_1-\left(j-\frac{1}{2}\right)\lambda_1\right)\right]\exp\left[{-\frac{\left(x_1-\left(j-\frac{1}{2}\right)\lambda_1\right)^2}{2\sigma^2}}\right],\, (j-1)\lambda_1 \le x_1 \le j\lambda_1,
    \label{eq:pres_gauss}
\end{equation}
where $A^+=0.51$ and $\omega^+ \approx 0.09$, set to be the coupled amplitude and frequency of case~10. The phase shift $\delta \phi$ is prescribed as the ensemble-averaged phase difference between adjacent panels in case~10, as discussed in a later section. This setup is designed to approximate the D-Psub dynamics while remaining in an open-loop configuration, in order to assess whether the blowing and suction spatial profile and the phase shift among D-Psubs are responsible for the observed frequency shift.

The resulting frequency response is shown in Fig.~\ref{fig:spectrum_39}(b). The panel-averaged pressure spectrum is concentrated around the forcing frequency, without any noticeable frequency shift. This result indicates that the observed frequency shift in the fully coupled system is not a consequence of the prescribed spatial forcing profile or the imposed phase difference between panels, but rather arises from the intrinsic fluid--structure interaction.

\begin{figure}
    \begin{overpic}[width=\textwidth]{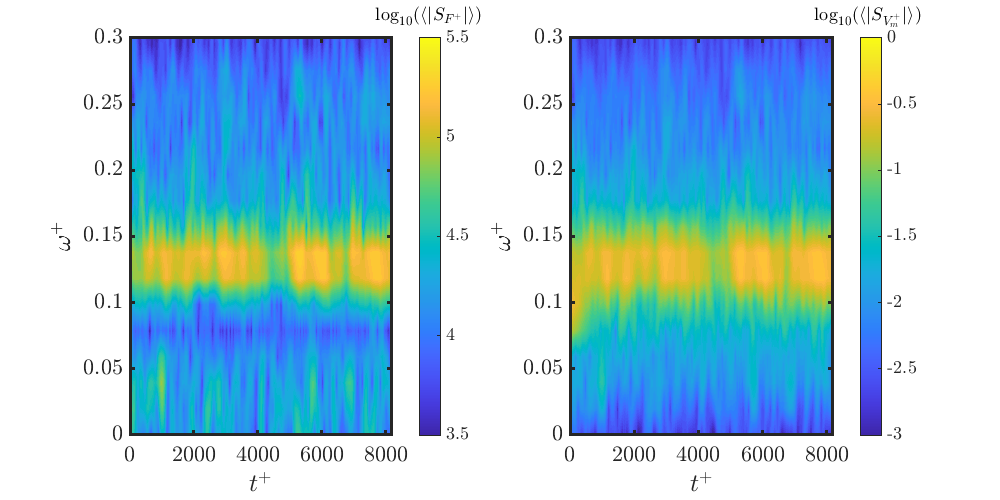}
            \put(14,48){\large\textbf{(a)}}  
            \put(56,48){\large\textbf{(b)}}  
    \end{overpic}
    \caption{Spectrogram of the (a) panel-averaged magnitude of the wall-normal force and (b) peak velocity for case~10, computed using a time window of $\Delta t^+ \approx 300$.}

    \label{fig:spectrogram_39}
\end{figure}

To investigate why the peak oscillation frequency does not align with the frequency of maximum game for a narrow-band phononic subsurface, we compute the spectrograms of both the input forcing and the resulting wall-normal velocity shown in Fig.~\ref{fig:spectrogram_39}. The spectrogram of a general time-dependent quantity $q(t)$ is defined as
\begin{equation}
S_{q}(t,\omega) = \int_{t}^{t+\Delta t} q(\tau) \exp\left({-\mathrm{i}\omega \tau}\right) d\tau,
\end{equation}
where $\Delta t$ denotes the temporal window required to resolve the dominant frequency content of $q$. By utilizing a sliding window of $\Delta t^+\approx 300$, we capture the transient spectral evolution while maintaining sufficient resolution of the governing dynamics.

The results indicate that during the initial stage, the surface velocity oscillates near the designed defect mode at $\omega^+\approx 0.1$. However, within $t^+ \approx 500$, the velocity spectrum shifts upward to an equilibrium coupled peak frequency of $\omega ^+\approx 0.13$. The underlying mechanism for this frequency shift is elucidated by the spectrogram of the wall-normal load, shown in Fig.~\ref{fig:spectrogram_39}(a). Although the surface velocity initially oscillates at the defect frequency, the wall pressure does not exhibit a corresponding response. Instead, the pressure forcing centers around the equilibrium peak frequency from the onset. This shift in the forcing spectrum drives the output velocity toward the equilibrium peak, even in regions where the system gain is sub-optimal.

In summary, although the mechanism underlying the frequency shift remains unclear, we have isolated it as a consequence of the coupled fluid--structure interaction. In particular, the pressure response is concentrated around the dominant frequency, even when the surface velocity oscillates at a different frequency, leading to the observed shift. This pressure-driven response cannot be reproduced by prescribed wall motion with fixed frequency and phase, indicating that the frequency shift is an intrinsic feature of the coupled system. The amplitude envelope at the defect frequency can therefore be regarded as a useful precursor for predicting the magnitude of the frequency shift. This indicates that the structural response is not solely dictated by the designed resonance, but is modified by the turbulent forcing, highlighting the bidirectional nature of the interaction even within a weakly coupled framework.

\section{Modification of turbulent flow statistics}\label{sec: turbulent stats}

In this section, we first examine the modification of turbulent drag via phononic subsurfaces. To further elucidate the underlying mechanisms responsible for drag reduction or increase, we analyze the one-point flow statistics together with the associated coherent structures. 

\subsection{Effect of phononic subsurfaces on skin friction}
\begin{figure*}
    \centering
    \includegraphics[width=0.5\textwidth]{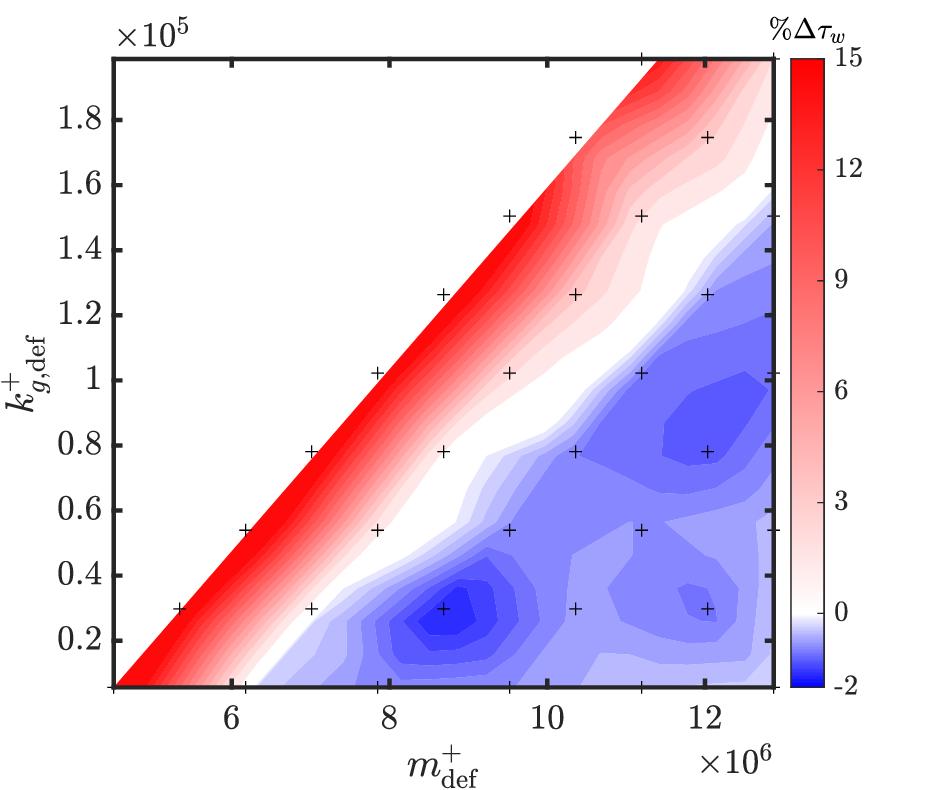}
    \caption{Contour map of the skin-friction change ratio $\%\Delta \tau_w$ as a function of the D-Psub parameters $(m^+_{\mathrm{def}}, k^+_{\mathrm{g,def}})$. The symbols ($+$) denote the parameter locations corresponding to the numerical experiments.}
    \label{fig:DR_map}
\end{figure*}
To quantify the change of skin friction, we first define the relative change in wall shear stress as
\begin{equation}
    \%\Delta \tau_w = \frac{\tau_{w,c}-\tau_{w,0}}{\tau_{w,0}}\times 100 \%,
\end{equation}
where $\tau_{w,c}$ denotes the wall shear stress of the controlled flow at the bottom wall, and $\tau_{w,0}$ is the mean wall shear stress of the uncontrolled flow. Positive values of $\%\Delta \tau_w$ indicate drag increase, whereas negative values correspond to drag reduction.

Fig.~\ref{fig:DR_map} presents the variation of skin-friction change across the D-Psub parameter space. The maximum drag reduction is observed at $(m^+_{\mathrm{def}}, k^+_{\mathrm{g,def}})=(8.6\times 10^6,\,2.9\times 10^4)$, yielding $\%\Delta \tau_w=-1.83$. As shown in Fig.~\ref{fig:DR_map}, the distribution of drag change closely follows the trend observed in the amplitude map in Fig.~\ref{fig:A_omega_map}(a). This behavior is consistent with prior studies of prescribed blowing and suction, where the drag response is primarily governed by the oscillation amplitude when the forcing frequency lies within a favorable range. In particular, for the present configuration with a fixed streamwise wavelength $\lambda_1^+ \approx 206$, forcing frequencies below $\omega^+ \approx 0.15$ in viscous units lead to drag reduction. In the current coupled system, the equilibrium oscillation frequency remains below $\omega^+ \approx 0.18$, and therefore the velocity amplitude becomes the dominant parameter controlling the drag response.

The corresponding coupled dynamics for the optimal drag reduction case (case~10) are $(\langle A_m^+\rangle, \lambda_1^+, \langle \omega^+ \rangle) = (0.514,\,206,\,0.127)$. This coupled response closely matches a data point from the prescribed blowing and suction dataset, $(A^+, \lambda^+, \omega^+) = (0.53,\,212.65,\,0.12)$, which yields $\%\Delta \tau_w = -1.89$ \citep{lin_24}. The close agreement between these results indicates that the prescribed forcing study provides a reliable reference for guiding the design of passive phononic subsurfaces.

For brevity, in the following analysis we consider two representative cases to investigate the effects of the D-Psub on the turbulent boundary layer. The first corresponds to the optimal drag-reduction case (case~10), and the second to a drag-increasing case (case~12). These two cases share the same defect mass $m_{\mathrm{def}}$ but differ in the defect stiffness $k_{\mathrm{g,def}}$.

\begin{figure}
    \centering
    \includegraphics[width=0.6\textwidth]{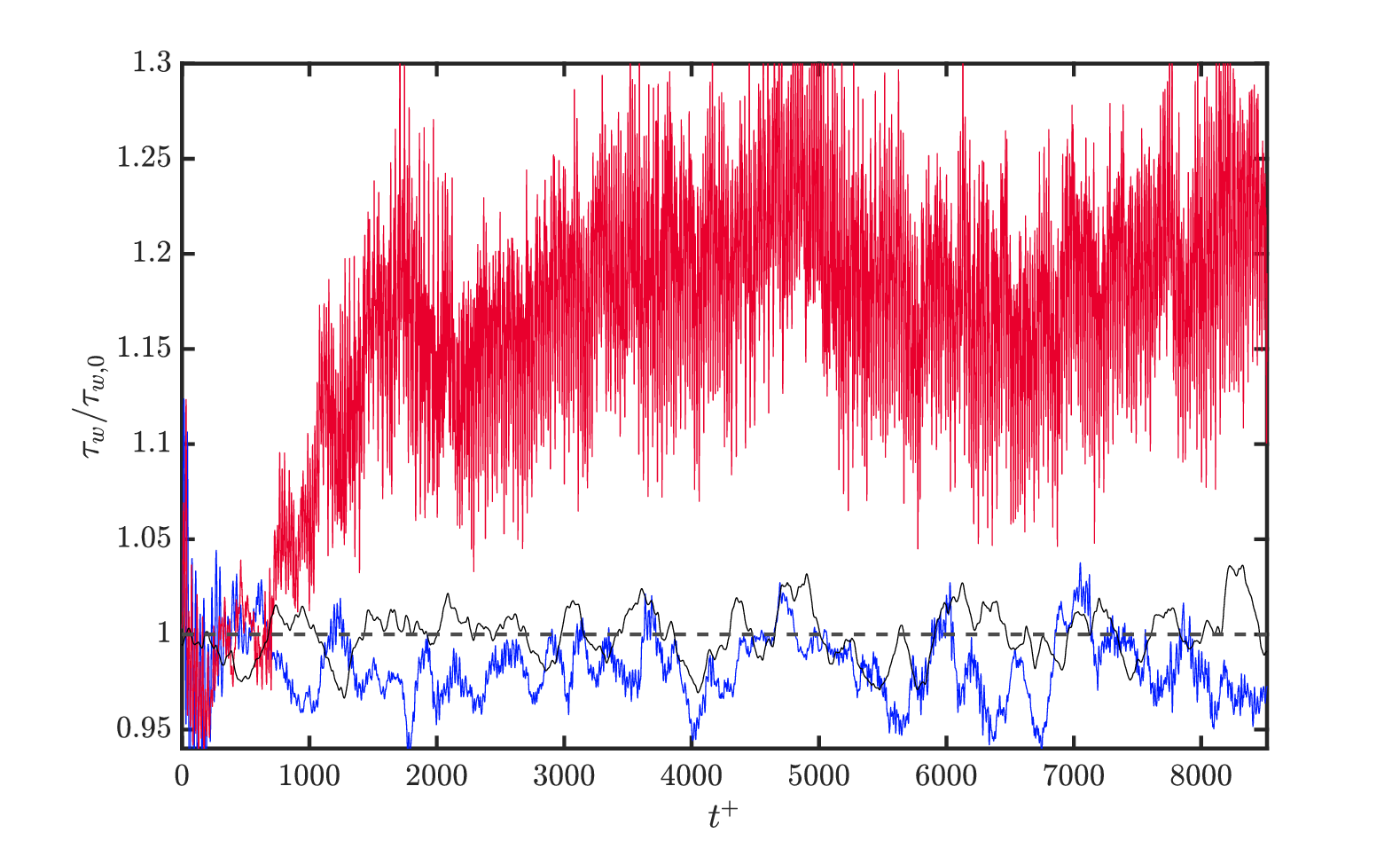}
    \caption{%
        Time evolution of wall-shear stress for case~10 (blue), case~12 (red), and uncontrolled base flow (black).
        Horizontal dash line indicates $\tau_w / \tau_{w,0}=1$.
    }
    \label{fig:drag_history_39_41}
\end{figure}

Fig.~\ref{fig:drag_history_39_41} shows the time series of wall shear stress for cases~10 and 12. In both cases, the D-Psub-controlled flow exhibits smaller temporal fluctuations in wall shear stress compared to corresponding prescribed blowing and suction cases in \citet{lin_24} (not shown). For the drag-reducing configuration (case~10), the fluctuation level is comparable to that of the uncontrolled flow, indicating that the passive D-Psub does not introduce additional variability in drag. In contrast, the drag-increasing case (case~12) exhibits larger fluctuations, consistent with stronger effective forcing at the wall. These results suggest that D-Psub-based control can reduce drag while maintaining low temporal variability in skin friction, which may be advantageous since large fluctuations are associated with intermittent high-drag events even when the mean drag is reduced.

\subsection{Effect of RPM subsurfaces on one-point statistics}
\begin{figure}
    \centering
    \includegraphics[width=\textwidth]{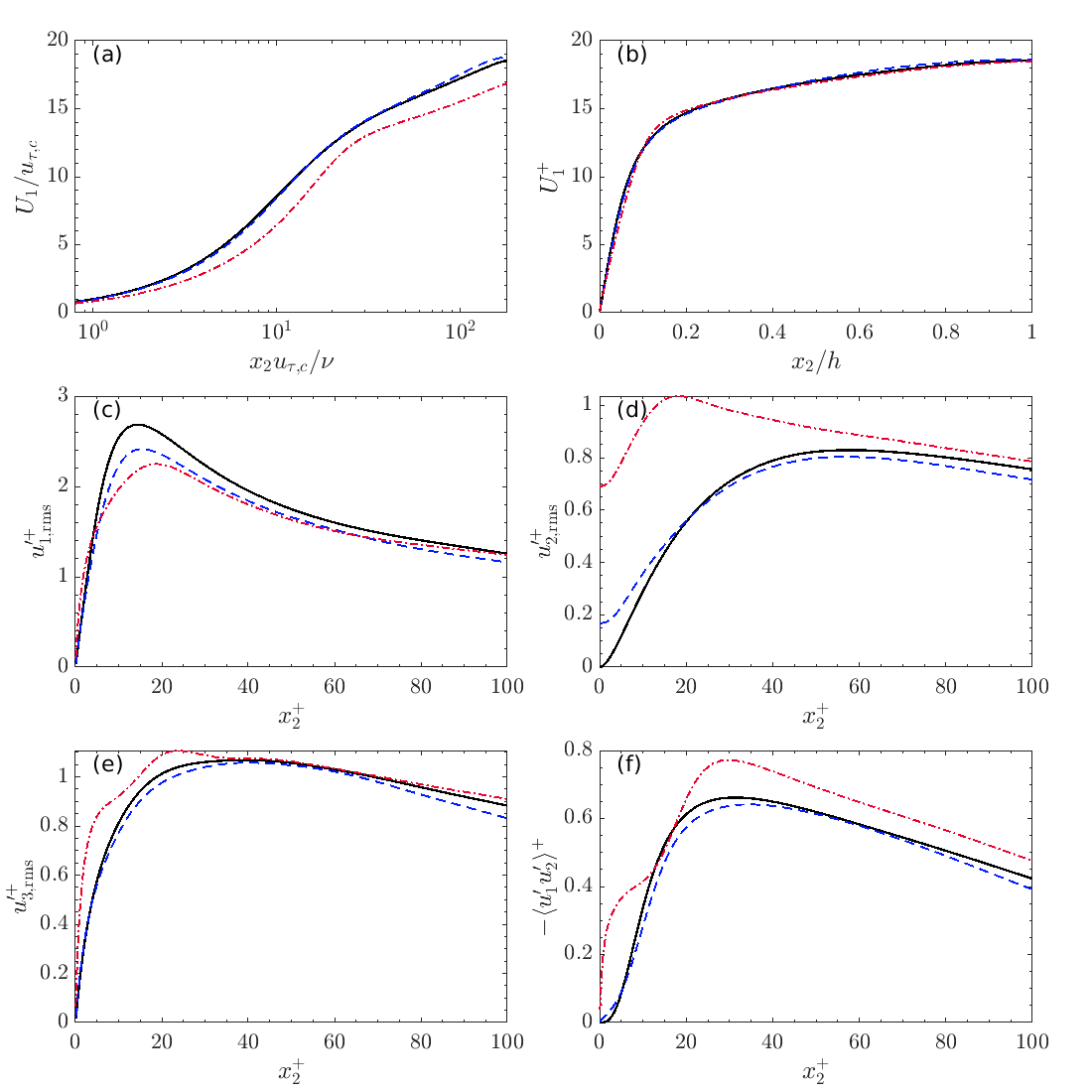}
    \caption{Mean streamwise velocity (a) normalized via friction velocity of controlled flow and (b)
    normalized via friction velocity of uncontrolled flow, root-mean-squared (c) streamwise, (d) wall-normal and (e) spanwise  velocity fluctuations, and (f) Reynolds shear stress for uncontrolled (black), case~10 (blue), and case~12 (red).}
    \label{fig:1D_stats}
\end{figure}
To understand the mechanisms governing drag modification induced by the D-Psub, we examine one-dimensional turbulence statistics in this section. The averaging operator is defined as an average over time and the two homogeneous directions, namely the streamwise and spanwise directions, while accounting for the out-of-phase dynamics of the D-Psubs. Fig.~\ref{fig:1D_stats} presents the mean streamwise velocity and associated turbulence statistics.

When normalized by the friction velocity of the controlled flow, as shown in Fig.~\ref{fig:1D_stats}(a), a clear velocity shift downward is observed for the drag-increasing case. In contrast, for the drag-reducing case, the velocity shift upward is less pronounced due to the relatively modest level of drag reduction. In Fig.~\ref{fig:1D_stats}(b), the mean streamwise velocity $U_1$ normalized by the uncontrolled flow quantities show a slightly flatter velocity profile away from the wall in the drag-increasing case, indicating enhanced turbulent mixing.

More pronounced differences are observed in the fluctuation statistics. As shown in Fig.~\ref{fig:1D_stats}(b), the root-mean-squared (rms) streamwise velocity fluctuation intensity $u_{1,\mathrm{rms}}^{\prime}$ is reduced in both controlled cases compared to the uncontrolled flow. This behavior is somewhat unexpected, as drag-increasing cases are typically associated with enhanced turbulence intensity. The observed reduction suggests a pumping effect induced by wall transpiration, which can be further understood through the TKE and pressure analyses presented in the next section. In particular, the local blowing phase generates a decelerating flow that suppresses streamwise velocity fluctuations in the buffer layer.

In contrast, the wall-normal fluctuation intensity shown in Fig.~\ref{fig:1D_stats}(c), exhibits a different trend: it is suppressed in the drag-reducing case but enhanced in the drag-increasing case, especially in the near-wall and buffer-layer regions. A similar behavior is observed for the spanwise fluctuation intensity. These results indicate that effective drag reduction is associated with the attenuation of both streamwise and wall-normal fluctuations, whereas excessive wall-normal motion enhances turbulence intensity and leads to drag increase. This phenomenon has also been observed in the breakdown of the drag-reduction regime of riblets, where the deterioration in performance is primarily attributed to the generation of spanwise rollers \citep{Garcia_Jimenez_2011}.

To further establish the connection between these observations and the drag response, we examine the Reynolds shear stress in the context of the  Fukagata-Iwamoto-Kasagi identity \citep{Fukagata_02}, which relates the skin-friction coefficient to the integral of the Reynolds shear stress across the channel. As shown in Fig.~\ref{fig:1D_stats}(d), the drag-reducing case exhibits a reduced magnitude of $-\langle u_1' u_2' \rangle$ compared to the uncontrolled flow. In contrast, the drag-increasing case shows an elevated Reynolds shear stress, particularly in the near-wall and buffer region. 

The wall-normal profile of Reynolds shear stress closely follows the behavior of the wall-normal velocity fluctuations, despite the general reduction observed in the streamwise fluctuations for both cases. This suggests that wall-normal velocity plays a dominant role in determining drag performance. In particular, controlling the magnitude of wall-normal motion is essential for achieving drag reduction, whereas excessive wall-normal forcing enhances Reynolds stress production and leads to drag increase.

\subsection{Effect of D-Psubs on turbulent kinetic energy}

In this section, we examine the turbulent kinetic energy (TKE) variation over a single D-Psub at different wall-normal locations, in order to shed light on the design of the spatial forcing profile for D-Psubs. However, due to quasi-periodic oscillations and phase differences across panels, direct time averaging would mask the response induced by the wall motion. To extract the turbulence modulation consistent with the panel dynamics, we introduce a phase-conditioned averaging procedure.

For each panel $j$, the interface velocity signal $V_{m,j}(t)$ is treated as the input signal $q(t)$ in Eq.~\eqref{eq:phase_hilbert}. The corresponding instantaneous phase is then defined as
\begin{equation}
    \phi_j(t)=\phi_{V_{m,j}}(t)+\frac{\pi}{2},
\end{equation}
where the phase shift provides a consistent physical interpretation: $\phi_j=0$ corresponds to a zero crossing, $\phi_j=\pi/2$ to the local maximum of the oscillation, and $\phi_j=3\pi/2$ to the local minimum.

The TKE is defined from the velocity fluctuations as
\begin{equation}
    k(x_1,x_2,x_3,t)=\frac{1}{2}\left(u_1'^2+u_2'^2+u_3'^2\right).
\end{equation}
We first average the TKE above the D-Psubs in the spanwise direction and over observed periods for each panel $j$, then average over all panels.
\begin{equation}
    K(\zeta_1,x_2,\phi)= \frac{1}{N_{\mathrm{DPsub}}} \sum_{j=1}^{N_{\mathrm{DPsub}}}\frac{1}{N_{\mathrm{per}}}\sum^{N_{\mathrm{per}}}_{p=1}\frac{1}{L_3}\int_0^{L_3} k((j-1)\lambda_1 + \zeta_1,x_2,x_3,\phi^p_j)\,\mathrm{d}x_3,
\end{equation}
where $0 \le \zeta_1 \le \lambda_1$, $N_{\mathrm{per}}$ is the number of periods identified from the evolution of the instantaneous phase and $\phi^p_j$ is the instantaneous phase of panel $j$ for the $p$-th period. The averaging is performed by grouping snapshots that correspond to the same phase, thereby preserving the phase-resolved dynamics of the flow. 

To quantify the effect of control, we evaluate the change in TKE relative to the uncontrolled flow,
\begin{equation}
    \Delta K= K_\mathrm{c}-K_\mathrm{0},
\end{equation}
where $K_\mathrm{c}$ and $K_\mathrm{0}$ denote the TKE of the controlled and uncontrolled flows, respectively.
\begin{figure}[t]
    \centering
    \includegraphics[width=\textwidth]{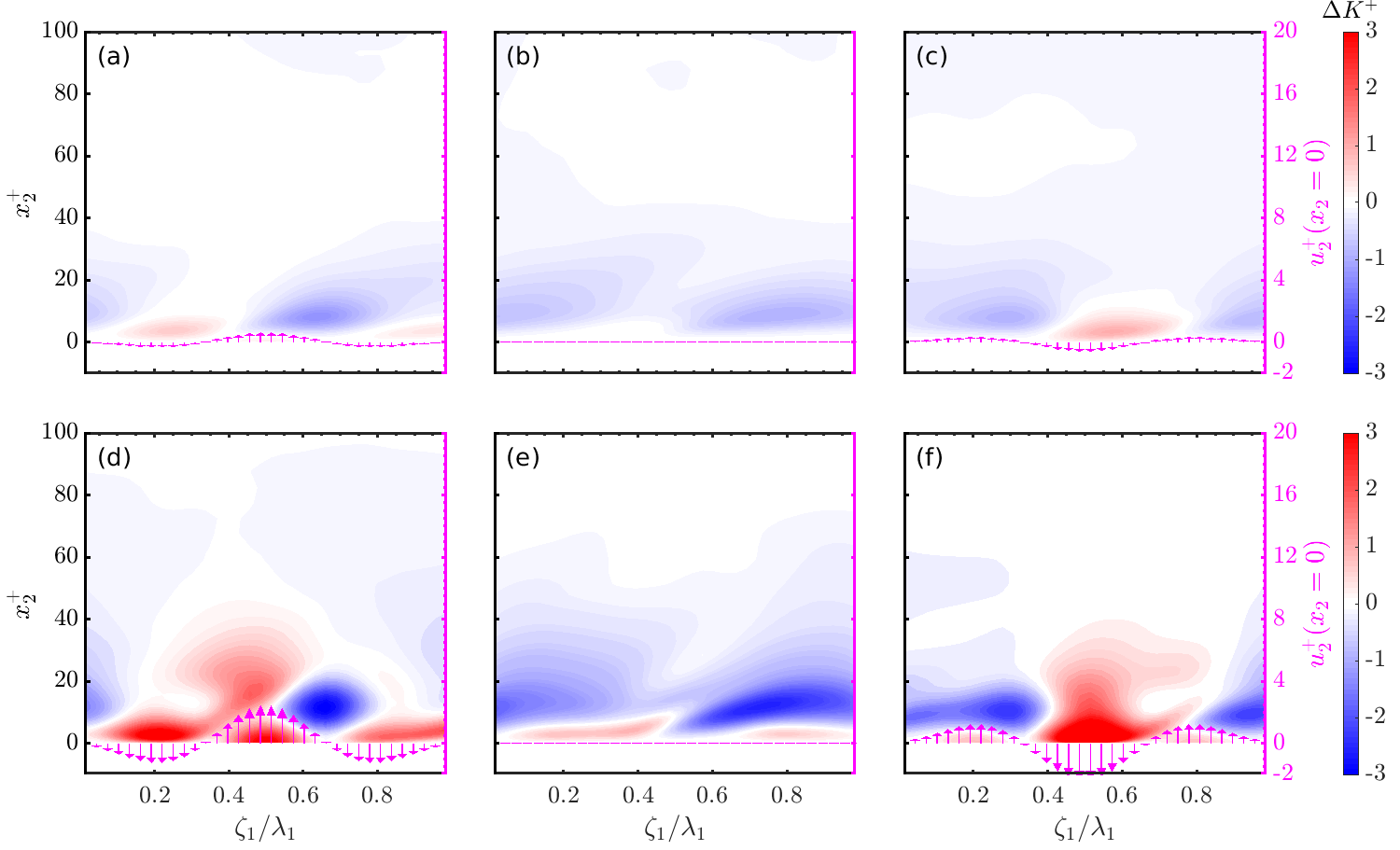}
    \caption{Phase-averaged TKE deviation $\Delta K$ for case~10 (top) and case~12 (bottom) for $\phi = 0.5\pi$~(a,d), $\pi$~(b,e), and $1.5\pi$~(c,f). The magenta arrows indicate the magnitude and direction of blowing and suction (right axis).
             }
    \label{fig:TKE_phase_comparison}
\end{figure}

Fig.~\ref{fig:TKE_phase_comparison} shows the contour maps of the TKE change $\Delta K$ for case~10 and case~12 as functions of the streamwise coordinate over a single panel $\zeta_1$ and wall-normal distance $x_2$. For the drag-reducing case (case~10), which is associated with relatively weak blowing and suction, the blowing phase leads to a reduction of TKE, while the suction phase results in a mild increase. In contrast, the drag-increasing case (case~12) exhibits a similar spatial pattern in the buffer region, but a markedly different response near the wall ($x_2^+ \leq 10$). In this region, both blowing and suction phases lead to an increase in TKE. This difference can be attributed to the stronger actuation amplitude in case~12. When the wall-normal velocity becomes sufficiently large, it directly enhances wall-normal velocity fluctuations, thereby increasing TKE in the near-wall region.

\begin{figure*}[t]
    \centering
    \includegraphics[width=\textwidth]{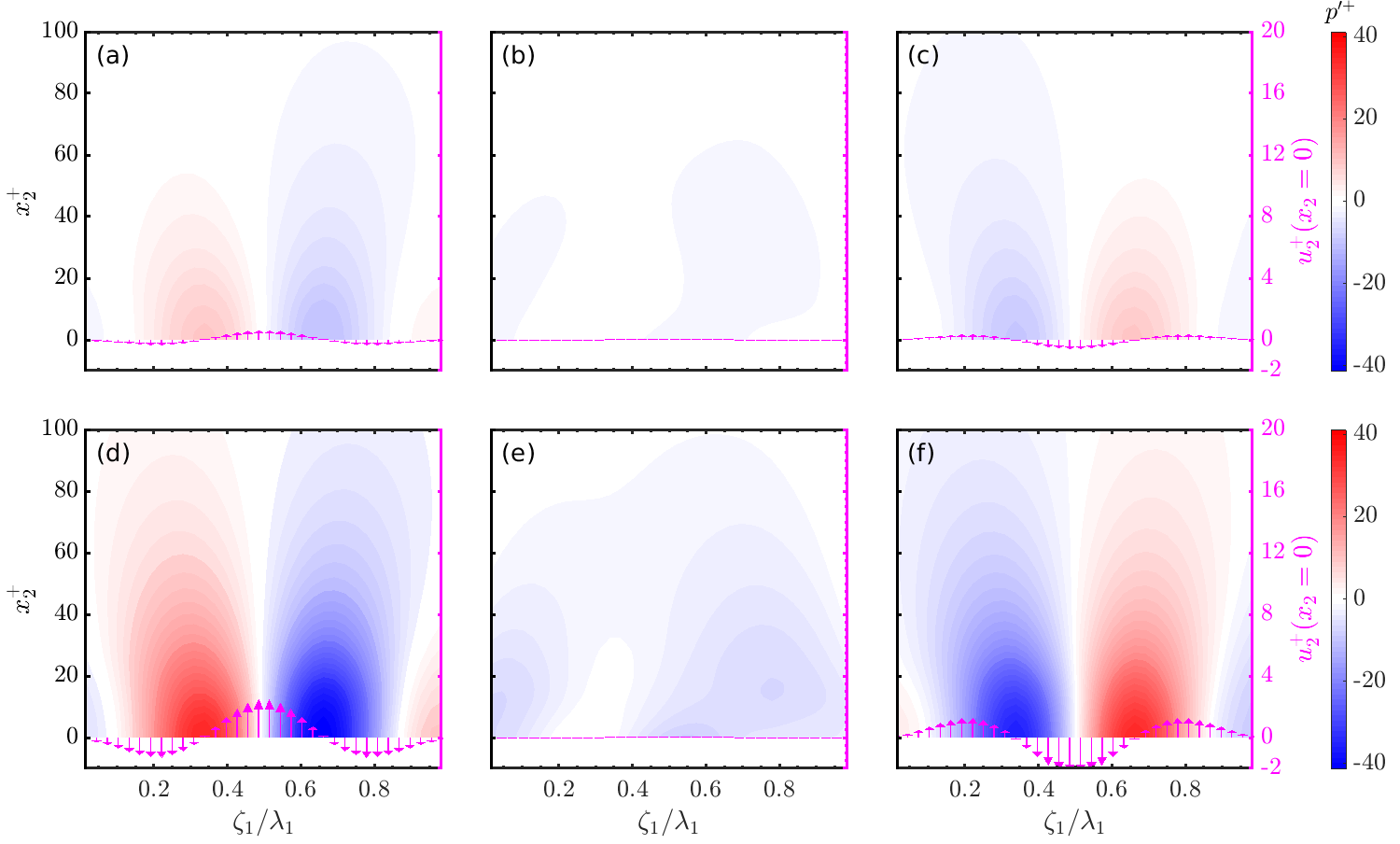}
    \caption{Phase-averaged pressure fluctuation $p^{\prime}$ for case~10 (top) and case~12 (bottom) for $\phi = 0.5\pi$~(a,d), $\pi$~(b,e), and $1.5\pi$~(c,f). The magenta arrows indicate the magnitude and direction of blowing and suction (right axis).}
    \label{fig:P_phase_comparison}
\end{figure*}
The phase-dependent change of TKE can be interpreted through the phase-averaged pressure fluctuation field. In Fig.~\ref{fig:P_phase_comparison}, the pressure fluctuation is averaged based on the instantaneous phase of the D-Psub wall motion, following the same phase-conditioned procedure used for TKE. The results show a favorable pressure gradient during the blowing phase ($\phi=\pi/2$) and an adverse pressure gradient during the suction phase ($\phi=3\pi/2$).

This behavior can be understood by considering the streamwise momentum equation evaluated at the wall,
\begin{equation}
   \frac{1}{\rho}\frac{\partial p}{\partial x_1}\Big|_{x_2=0}
    =
    -u_2\frac{\partial u_1}{\partial x_2}\Big|_{x_2=0}
    +\nu\frac{\partial^2 u_1}{\partial x_2^2}\Big|_{x_2=0}.
\end{equation}
Under the imposed mean pressure gradient in channel flow, the mean shear terms, ${\partial u_1}/{\partial x_2}$ and ${\partial^2 u_1}/({\partial x_2)^2}$ are positive. As a result, the sign of the pressure gradient fluctuation is determined by the wall-normal velocity $u_2$ at the wall. During the blowing phase ($u_2>0$), a relative favorable pressure gradient ($\partial p/\partial x_1<0$) is induced, which accelerates the near-wall flow. In contrast, during the suction phase ($u_2<0$), a relative adverse pressure gradient is generated, leading to local flow fluctuation deceleration. This phase-dependent behavior is observed consistently across different actuation amplitudes, as shown in Fig.~\ref{fig:P_phase_comparison}.

These pressure-gradient effects provide a direct explanation for the observed modification of the TKE. During the blowing phase, the local favorable pressure gradient accelerates the streamwise flow and reduces velocity fluctuations, analogous to relaminarization. In contrast, during the suction phase, the induced adverse pressure gradient decelerates the streamwise flow and enhances velocity fluctuations, analogous to transition. This mechanism is consistent with oscillatory channel flow, where \citet{Ebadi_White_Pond_Dubief_2019} reported that the streamwise velocity fluctuation reaches a minimum during the accelerating phase and a maximum during the decelerating phase of the mean pressure gradient.

This behavior is observed in both drag-reducing and drag-increasing cases within the buffer layer. However, the difference in drag response arises from the near-wall region. Close to the wall, the wall-normal velocity induced by the control input can become comparable to, or exceed, the streamwise velocity due to the no-slip boundary condition. In the drag-reducing case, the control amplitude is relatively small, and the TKE increase remains confined to a very thin layer near the wall. This increase is rapidly compensated by the reduction of streamwise velocity fluctuations away from the wall. 

In contrast, the drag-increasing case exhibits a pronounced increase in TKE across the near-wall region, extending into the buffer layer. This is caused by excessive blowing and suction, which directly amplify wall-normal velocity fluctuations. As a result, the increase in turbulence intensity outweighs the reduction in streamwise fluctuations induced by the favorable pressure gradient. This combined effect explains the trends observed in the one-point statistics: while streamwise fluctuations are reduced in both cases, the wall-normal and spanwise fluctuations decrease in the drag-reducing case but increase in the drag-increasing case.

\subsection{Effects of phononic subsurfaces on coherent structures}

\begin{figure*}[t]
    \centering
    \begin{minipage}[t]{0.49\textwidth}
        \hfill
        \begin{overpic}[width=\textwidth]{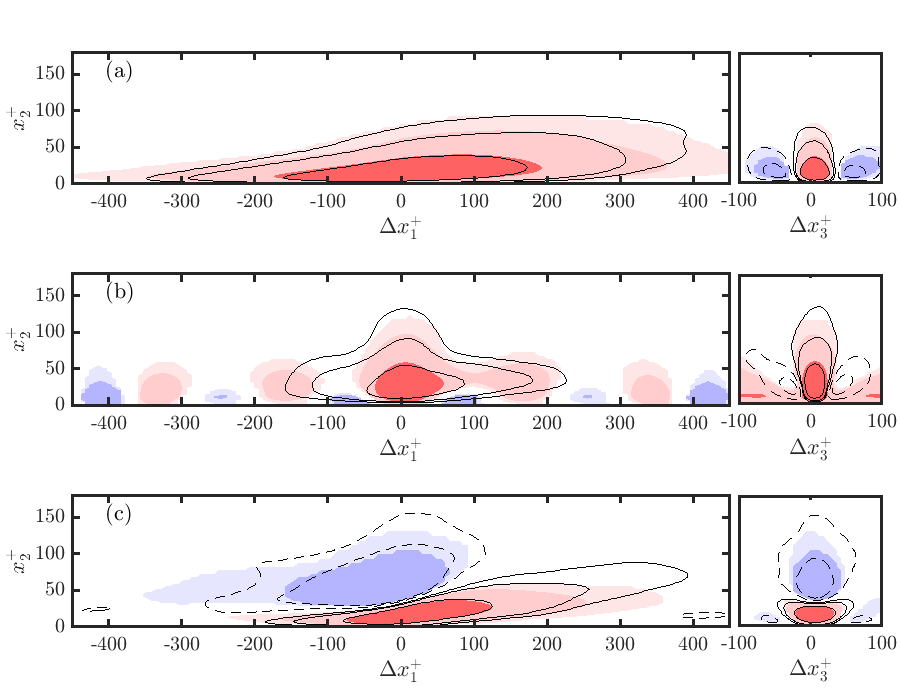}
        \end{overpic}
    \end{minipage}
    \hfill
    \begin{minipage}[t]{0.49\textwidth}
        \hfill
        \begin{overpic}[width=\textwidth]{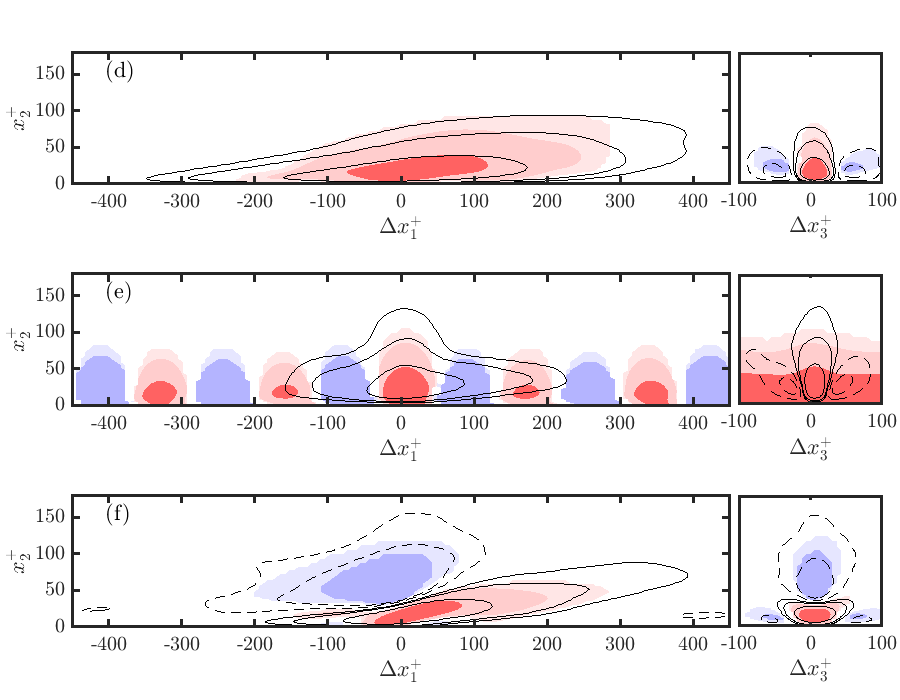}
        \end{overpic}
    \end{minipage}
    \caption{Two-point correlation function $C_{11}$ (a,d), $C_{22}$ (b,e), and $C_{33}$ (c,f) shown in the $\Delta x_1^+$--$x_2^+$ plane 
    (left panels) and $\Delta x_3^+$--$x_2^+$ plane (right panels) for case~10 (a--c) and 
    case~12 (d--f). Contours levels are $[-0.15, -0.05, 0.05, 0.10, 0.30]$ for the controlled (color) and uncontrolled flow (lines), where red/solid lines indicates positive levels and blue/dashed lines denote negative levels.}
    \label{fig:two_point_correlation_both}
\end{figure*}

Two-point correlation functions are computed to examine how coherent structures are modified in the presence of phononic subsurfaces. The two-point velocity correlation is defined as
\begin{equation}
    C_{ij}(\Delta x_1,\Delta x_3,x_2;x_{2,\text{ref}})=\frac{\langle u'_i(x_1,x_2,x_3)u'_j(x_1+\Delta x_1,x_{2,\text{ref}},x_3+\Delta x_3\rangle}{\langle u'_i(x_1,x_2,x_3)u'_j(x_1,x_{2,\text{ref}},x_3)\rangle},
\end{equation}
where $u'_i$ denotes the velocity fluctuation obtained by subtracting the time-averaged mean. Owing to homogeneity in the streamwise and spanwise directions, together with statistical stationarity, $C_{ij}$ in channel flow depends on $\Delta x_1$, $\Delta x_3$, $x_2$, and the reference wall-normal location $x_{2,\text{ref}}$. In the present study, $x^+_{2,\text{ref}}=15$ is chosen to focus on the near-wall region where turbulence production is most active. The mid-plane cross sections of the correlation function are shown in Fig.~\ref{fig:two_point_correlation_both}.

In the streamwise autocorrelation $C_{11}$, the drag-reducing case demonstrates elongated and attached streaks compared to the uncontrolled flow. In contrast, the drag-increasing case shows compressed streaks with a similar inclination angle. The attached streaks in the drag-reducing case suggest that the control input suppresses streak breakdown, allowing the structures to persist and extend downstream through advection. This behavior is consistent with attenuation of the lift-up mechanism, which limits the amplification of streamwise velocity fluctuations and contributes to drag reduction.

The wall-normal autocorrelation $C_{22}$ is strongly influenced by the imposed blowing and suction, as evidenced by the periodic patterns observed in the $\Delta x_1$--$x_2$ plane. This effect is more pronounced in the drag-increasing case due to the larger actuation amplitude. Additional insight into the quasi-streamwise vortices structures can be obtained from the spanwise autocorrelation $C_{33}$. In the $\Delta x_3$--$x_2$ plane, both cases exhibit a reduction in the spanwise extent of the quasi-streamwise vortices, indicating a general weakening of these structures. However, in the $\Delta x_1$--$x_2$ plane, the drag-reducing case shows elongated and more coherent vortices, whereas the drag-increasing case exhibits compressed structures. These trends are consistent with our previous observations in prescribed blowing and suction studies~\citep{lin_24}.

Overall, the results indicate that quasi-streamwise vortices are weakened in both cases, which tends to suppress turbulence production. However, unlike other control strategies such as riblets or permeable surfaces, where drag increase is often associated with the generation of spanwise rollers~\citep{Garcia_Jimenez_2011}, the drag-increasing mechanism observed here is different. In the present configuration, drag increase primarily arises from strong wall-normal velocity fluctuations near the wall, which enhance turbulent momentum transfer and generate additional Reynolds shear stress due to the control input.

\section{Conclusion}\label{sec: conclusion}

In this study, we investigated the weakly coupled fluid--structure interaction between turbulent channel flow and defect-embedded phononic subsurfaces (D-Psubs). The results demonstrate that D-Psubs can passively respond to broadband turbulent forcing and, in turn, feed back onto the flow in a structured and dynamically consistent manner. We observe measurable drag reduction in selected configurations, although, the primary outcome of this work is the characterization of the coupled dynamics governing turbulence--phononic subsurface interaction.

From the perspective of structural dynamics, the coupled system exhibits behavior that cannot be inferred from uncoupled models alone. In particular, the dominant oscillation frequency shifts away from the designed defect resonance and scales with the amplitude envelope, highlighting the role of fluid--structure interaction in determining the effective response of the system. In addition, the phase relationship between adjacent panels is governed by the convection velocity of turbulent structures, indicating that the structural response is closely tied to the advective dynamics of the flow. These findings demonstrate that the coupled behavior emerges from the interplay between structural resonance and turbulent transport, rather than being prescribed a priori.

The results further identify three key quantities—the coupled equilibrium amplitude, the effective oscillation frequency, and the amplitude envelope—as fundamental descriptors of the D-Psub dynamics. While the latter two can be characterized from structural considerations, the coupled amplitude arises from the interaction with turbulence and cannot be predicted from the structural model alone. Developing predictive models for this quantity would enable more efficient design of dynamically responsive surfaces without reliance on high-fidelity simulations.

The interaction between the D-Psubs and the flow leads to systematic modifications of near-wall turbulence. Phase-averaged analysis reveals that wall-normal motion induces alternating regions of turbulence attenuation and amplification, associated with local favorable and adverse pressure gradients. These effects manifest in changes to coherent structures, including elongation and increased attachment of streamwise streaks and attenuation of quasi-streamwise vortices. However, similar structural modifications are also observed in drag-increasing cases, indicating that the net drag response is not determined solely by the suppression of coherent structures, but rather by the balance between induced wall-normal motions and the resulting Reynolds stresses.

Overall, this work establishes a weakly coupled framework for studying turbulence--phononic subsurface interaction and demonstrates that resonant subsurfaces can selectively respond to and reorganize turbulent flow structures. These findings provide a foundation for the design of dynamic passive surfaces that interact with specific turbulent scales through frequency-selective feedback. Future work should investigate the limitations of the weakly coupled formulation and extend the approach to fully coupled models that account for both displacement and velocity interactions between the subsurface and the flow.

\section*{Acknowledgments}
This material is based upon work supported by the Air Force Office of Scientific Research under award number FA9550-23-1-0299. 

%
%
%
%
%
%

\appendix
\section{Complete Parameter Space of D-Psub Configurations}
\label{sec:appendix_RPM}

For completeness, we provide the full set of defect-embedded phononic subsurface (D-Psub)configurations considered in the present study. The table includes all combinations of defect mass and stiffness explored in the parametric sweep, together with the corresponding structural response and flow statistics.

The defect parameters are characterized by mass and stiffness, $(m_{\mathrm{def}}, k_{g,\mathrm{def}})$. In addition, we report the effective excitation amplitude $A_E$ and the corresponding defect frequency $\omega_{\mathrm{def}}$, which characterize the intrinsic response of the D-Psub. The coupled flow--structure interaction is quantified through the resulting oscillation amplitude $A_m$, the peak response frequency $\omega^{*}$, and the percentage change in wall shear stress, $\%\Delta \tau_w$. The complete dataset, comprising 34 cases, is summarized in Table~\ref{tab:RPM_cases}. 

\begin{table}
\centering
\caption{D-Psub defect parameters and coupled flow--structure response metrics.}
\label{tab:RPM_cases}
\renewcommand{\arraystretch}{1.5}
\begin{tabular}{c c c c c c c c || c c c c c c c c}
\hline\hline
\multicolumn{8}{c||}{\textbf{Cases 1--17}} &
\multicolumn{8}{c}{\textbf{Cases 18--34}} \\
\cline{1-16}
ID &
$m_{\mathrm{def}}^{+}\times 10^{-6}$ &
$k_{g,\mathrm{def}}^{+}\times 10^{-4}$ &
$A_E^{+}$ &
$\omega_{\mathrm{def}}^{+}$ &
$A_m^{+}$ &
$\omega^{*+}$ &
$\%\Delta \tau_w$ &
ID &
$m_{\mathrm{def}}^{+}\times 10^{-6}$ &
$k_{g,\mathrm{def}}^{+}\times 10^{-4}$ &
$A_E^{+}$ &
$\omega_{\mathrm{def}}^{+}$ &
$A_m^{+}$ &
$\omega^{*+}$ &
$\%\Delta \tau_w$ \\
\hline

1  & 4.504 & 0.568 & 1.170 & 0.1155 & 2.286 & 0.1845 & 19.07 &
18 & 11.197 & 10.222 & 0.471 & 0.1183 & 0.480 & 0.1397 & -1.00 \\
2  & 5.341 & 2.981 & 0.987 & 0.1256 & 1.994 & 0.1737 & 17.05 &
19 & 11.197 & 15.049 & 0.471 & 0.1353 & 1.292 & 0.1576 & 1.92 \\
3  & 6.178 & 0.568 & 0.854 & 0.0987 & 0.936 & 0.1424 & -0.19 &
20 & 11.197 & 19.876 & 0.471 & 0.1504 & 2.155 & 0.1709 & 15.92 \\
4  & 6.178 & 5.395 & 0.854 & 0.1325 & 2.230 & 0.1731 & 19.10 &
21 & 12.033 & 2.981 & 0.439 & 0.0837 & 0.426 & 0.0985 & -1.03 \\
5  & 7.014 & 2.981 & 0.752 & 0.1096 & 0.929 & 0.1469 & -0.12 &
22 & 12.033 & 7.808 & 0.439 & 0.1050 & 0.679 & 0.1254 & -1.22 \\
6  & 7.014 & 7.808 & 0.752 & 0.1375 & 2.165 & 0.1728 & 17.99 &
23 & 12.033 & 12.635 & 0.438 & 0.1226 & 0.926 & 0.1433 & -0.74 \\
7  & 7.851 & 0.568 & 0.672 & 0.0876 & 0.581 & 0.1218 & -0.85 &
24 & 12.033 & 17.462 & 0.438 & 0.1380 & 1.409 & 0.1585 & 3.62 \\
8  & 7.851 & 5.395 & 0.672 & 0.1175 & 1.238 & 0.1505 & 1.04 &
25 & 12.870 & 0.568 & 0.410 & 0.0684 & 0.287 & 0.0842 & -0.30 \\
9  & 7.851 & 10.222 & 0.672 & 0.1413 & 2.243 & 0.1731 & 17.31 &
26 & 12.870 & 5.395 & 0.410 & 0.0918 & 0.373 & 0.1111 & -0.46 \\
10 & 8.687 & 2.981 & 0.607 & 0.0985 & 0.514 & 0.1272 & -1.83 &
27 & 12.870 & 10.222 & 0.410 & 0.1104 & 0.633 & 0.1299 & -1.15 \\
11 & 8.687 & 7.808 & 0.607 & 0.1235 & 1.117 & 0.1522 & 0.47 &
28 & 12.870 & 15.049 & 0.410 & 0.1262 & 1.011 & 0.1451 & -0.55 \\
12 & 8.687 & 12.635 & 0.607 & 0.1443 & 2.245 & 0.1720 & 18.28 &
29 & 12.870 & 19.876 & 0.410 & 0.1403 & 1.289 & 0.1594 & 2.31 \\
13 & 9.524 & 0.568 & 0.554 & 0.0795 & 0.412 & 0.0994 & -0.74 &
30 & 10.360 & 2.981 & 0.509 & 0.0902 & 0.419 & 0.1075 & -0.69 \\
14 & 9.524 & 5.395 & 0.554 & 0.1067 & 0.707 & 0.1334 & -0.78 &
31 & 10.360 & 7.808 & 0.509 & 0.1131 & 0.862 & 0.1370 & -0.91 \\
15 & 9.524 & 10.222 & 0.554 & 0.1283 & 1.410 & 0.1540 & 2.00 &
32 & 10.360 & 12.635 & 0.509 & 0.1321 & 1.335 & 0.1558 & 3.31 \\
16 & 9.524 & 15.049 & 0.554 & 0.1467 & 2.331 & 0.1731 & 17.46 &
33 & 10.360 & 17.462 & 0.509 & 0.1487 & 1.885 & 0.1764 & 10.31 \\
17 & 11.197 & 0.568 & 0.471 & 0.0734 & 0.436 & 0.0887 & -0.37 &
34 & 12.033 & 17.462 & 0.438 & 0.1380 & 1.409 & 0.1585 & 3.62 \\

\hline\hline
\end{tabular}
\end{table}

\newpage

\bibliography{Ref}

\end{document}